\documentclass[preprint,12pt]{elsarticle}

\usepackage[dvipsnames]{xcolor}

\usepackage{graphicx,dcolumn,bm,microtype,hyperref,multirow,amscd,amsmath,amssymb,amsfonts,physics}
\usepackage{url,booktabs,nicefrac,lipsum,graphicx,float,placeins,tikz}
\usepackage[utf8]{inputenc}
\usepackage[T1]{fontenc}
\usepackage[version=4]{mhchem}
\usepackage{algorithmicx,algorithm,algpseudocode}
\usepackage{ulem}

\usepackage{multirow}
\usepackage{subcaption}
\usepackage{pgfplots}
\pgfplotsset{compat=1.18} 
\usetikzlibrary{pgfplots.groupplots}
\usepackage{xcolor}

\definecolor{slateblue}{rgb}{0.4157, 0.3529, 0.8039}
\definecolor{lightblue}{rgb}{0.50196, 0.76863, 0.91373}
\definecolor{midnightblue}{RGB}{25,25,112}
\definecolor{cornflowerblue}{RGB}{100, 149, 237}
\definecolor{lavender}{RGB}{230, 230, 250}
\definecolor{powderblue}{rgb}{0.69, 0.88, 0.9}
\definecolor{paleturquoise1}{rgb}{0.69, 0.93, 0.93}
\definecolor{lightcyan}{rgb}{0.88, 1.0, 1.0}
\definecolor{mintcream}{rgb}{0.96, 1.0, 0.98}
\definecolor{beige}{rgb}{0.96, 0.96, 0.86}
\definecolor{antiquewhite}{rgb}{1, 0.972, 0.953}
\definecolor{silver}{rgb}{0.459, 0.525, 0.580}
\definecolor{silverblue}{rgb}{0.251, 0.365, 0.447}
\definecolor{navyblue}{rgb}{0.0, 0.0, 0.5}
\definecolor{customgrey}{rgb}{0.9451, 0.9451, 0.9451}
\definecolor{lightmaroon}{rgb}{0.7804, 0.7176, 0.6392} 
\definecolor{wheat}{rgb}{0.9176, 0.8588, 0.7843} 
\definecolor{lightolive}{rgb}{0.7098, 0.7569, 0.5569}
\definecolor{olivedrab}{rgb}{0.3098, 0.4353, 0.3216} 
\definecolor{paleturquoise}{rgb}{0.498, 0.827, 0.690}
\definecolor{azure}{rgb}{0.7843, 1.0, 0.8784}
\definecolor{lightpink}{rgb}{1.0, 0.8196, 0.8902}
\definecolor{lightpurple}{rgb}{0.7490, 0.6745, 0.8863}
\definecolor{purple}{rgb}{0.6275, 0.5176, 0.8627}
\definecolor{darkpurple}{rgb}{0.3922, 0.3608, 0.7333}
\definecolor{bisk}{rgb}{1.0, 0.9843, 0.9608}
\definecolor{violet}{rgb}{0.7412, 0.5137, 0.8078} 
\definecolor{lightviolet}{rgb}{0.8980, 0.6902, 0.9176}
\definecolor{powderpink}{rgb}{0.9451, 0.7765, 0.9059} 
\definecolor{lightpowderpink}{rgb}{0.9686, 0.9098, 0.9647}
\definecolor{seashell}{rgb}{0.9765, 0.9765, 0.9765}

\makeatletter
\newcommand\resetstackedplots{
\makeatletter
\pgfplots@stacked@isfirstplottrue
\makeatother
\addplot [forget plot,draw=none] coordinates{(1,0) (2,0) (3,0)};
}
\makeatother

\newcolumntype{H}{>{\setbox0=\hbox\bgroup}c<{\egroup}@{}}

\algnewcommand\algorithmicswitch{\textbf{switch}}
\algnewcommand\algorithmiccase{\textbf{case}}
\algnewcommand\algorithmicassert{\texttt{assert}}
\algnewcommand\Assert[1]{\State \algorithmicassert(#1)}
\algdef{SE}[SWITCH]{Switch}{EndSwitch}[1]{\algorithmicswitch\ #1\
\algorithmicdo}   {\algorithmicend\ \algorithmicswitch}
\algdef{SE}[CASE]{Case}{EndCase}[1]{\algorithmiccase\ 
#1}{\algorithmicend\  \algorithmiccase} \algtext*{EndSwitch}
\algtext*{EndCase}
\algrenewcommand{\algorithmiccomment}[1]{$\triangleright$
\textit{#1}}
\algnewcommand{\algorithmicgoto}{\textbf{go to}}%
\algnewcommand{\Goto}[1]{\algorithmicgoto~\ref{#1}}%

\biboptions{numbers,sort&compress}

\hypersetup{pdfauthor={Name}}

\journal{Advances in Quantum Chemistry}

\begin{document}

\begin{frontmatter}

\title{Enhancing the Computational Efficiency of the DoNOF Program through a New Orbital Sorting Scheme}

\author{Élodie Boutou$^{1,2,3}$}
\ead{elodie.boutou@etu.uca.fr}
\author{Juan Felipe Huan Lew-Yee$^{3}$}
\ead{felipe.lew.yee@dipc.org}
\author{Jose Maria Mercero$^{1,3}$}
\ead{jm.mercero@ehu.eus}
\author{Mario Piris$^{1,3,4}$}
\ead{mario.piris@ehu.eus}

\address{$^{1}$Kimika Fakultatea, Euskal Herriko Unibertsitatea (UPV/EHU), 20018 Donostia, Spain.}
\address{$^{2}$Polytech Clermont, University of Clermont Auvergne, 63178 Aubière Cedex, France.}
\address{$^{3}$Donostia International Physics Center (DIPC), 20018 Donostia, Spain.}
\address{$^{4}$Basque Foundation for Science (IKERBASQUE), 48013 Bilbao, Spain.}

\begin{abstract}
This work presents a novel approach to distribute orbitals into subspaces within electron-pairing-based natural orbital functionals (NOFs). This approach modifies the coupling between weakly and strongly occupied orbitals by applying an alternating orbital sorting strategy. In contrast to the previous orbital sorting that enforced electron pairing within subspaces of contiguous orbitals, the new approach provides greater flexibility, enabling a calculation scheme where the size of the subspaces can be gradually expanded. As a consequence, one can start using subspaces of only one weakly occupied orbital (perfect pairing) and progressively enlarge their size by incorporating more weakly occupied orbitals (extended pairing) up to the maximum size allowed by the basis set. In this way, the alternate orbital sorting allows solving first a simpler problem with small subspaces and leverage its orbital solution for the more intensive problem with larger subspaces, thereby reducing the overall computational cost and improving convergence, as we observed in the DoNOF program. The efficiency provided by the new sorting approach has been validated through benchmark calculations in \ce{H2O}, \ce{H2O2}, and \ce{NH3}. In particular, we compared three strategies: i) solving directly the calculation with the largest subspaces (one-shot strategy), as was usually done before this work, ii) starting with perfect pairing and stepwise increasing the number of orbitals in the subspaces one by one until reaching the maximum size (incremental strategy), and iii) starting with perfect pairing and transitioning directly to the maximum subspace size (two-step strategy). Our results show that the two-step approach emerges as the most effective strategy, achieving the lowest computational cost while maintaining high accuracy. These results confirm that the alternating orbital sorting scheme provides a robust and scalable framework for improving NOF calculations and could be particularly advantageous for extending these methods to larger and strongly correlated systems.
\end{abstract}


\end{frontmatter}

\section{\label{sec:Intro}Introduction}

In recent years, Natural Orbital Functional Theory (NOFT) \cite{Gilbert1975, Donnelly1978, Donnelly1979, Levy1979, Valone1980} has emerged as a powerful framework to describe electron correlation \cite{Piris2024a}. Unlike traditional wavefunction-based approaches, which often suffer from steep computational scaling, NOFT provides a formally fifth-order scaling, which can be reduced to fourth-order \cite{Lew-Yee2021, Lemke2022, Lew-Yee2023a}, making it significantly more efficient while maintaining an accurate description of correlated electronic states. This efficiency advantage is particularly relevant for multireference systems, where NOFT provides deeper insights than density functional approximations \cite{Lopez2011, Ruiperez2013, Ramos-Cordoba2015, Mitxelena2017b, Lopez2019, Mitxelena2020a, Mitxelena2020b, Mitxelena2022, Mitxelena2024}. By circumventing the limitations of conventional multireference methods, which rely on small active spaces and quickly become computationally intractable, NOFT enables the treatment of larger correlated electron systems while maintaining accuracy.

Beyond its success in ground-state electronic structure calculations, NOFT has been extended to describe excited states and molecular dynamics, broadening its applicability to time-dependent phenomena. Recent developments have demonstrated that NOFT can accurately capture excited-state properties by coupling natural orbital functionals with the extended random phase approximation (ERPA) \cite{Chatterjee2012, Lew-Yee2024}. These advances enable a systematic treatment of excited-state correlation effects, offering an alternative to traditional time-dependent density functional theory (TDDFT) or multireference wavefunction methods, particularly for strongly correlated systems. In addition, the application of NOFT to ab initio molecular dynamics (AIMD) has provided new insights into electronic structure evolution in real-time simulations, demonstrating its ability to describe dynamic electron correlation effects \cite{RiveroSantamaria2024, Piris2024c}. Furthermore, NOFT has shown remarkable improvements in addressing delocalization errors, a common challenge in density functional theory (DFT) that impacts the accuracy of charge transfer and reaction energy calculations \cite{Lew-Yee2023b}.

Despite these advances, further optimization of NOFT methods is necessary to enhance computational performance and extend applicability to more complex systems. A key development in this direction has been the incorporation of modern numerical techniques inspired by deep learning \cite{Lew-Yee2025}, which have significantly improved the convergence behavior of NOF calculations. In particular, the introduction of momentum-based techniques such as the ADAM optimizer for natural orbital optimization has led to substantial speed-ups in reaching self-consistency. Reference \cite{Lew-Yee2025} demonstrated the impact of these advances by presenting the largest NOF calculations to date, including a 1000-electron hydrogen cluster, a system that would be computationally prohibitive for conventional wavefunction-based methods. By representing the electronic structure in terms of natural orbitals \cite{Lowdin1955} and their corresponding occupation numbers, NOFT provides a distinct and comprehensive perspective on electron correlation, leading to the development of various functionals with different correlation treatments, as reviewed in \cite{Piris2007ch, Pernal2016, Schade2017, Piris2024b}.

Among these developments, Piris natural orbital functionals (PNOFs) \cite{Piris2006, Piris2007, Piris2010, Piris2010a, Piris2011, Piris2013a, Piris2014a, Piris2017, Mitxelena2018, Piris2021} have emerged as a highly effective realization of NOFT, offering a balance between accuracy and computational efficiency. These functionals explicitly incorporate electron correlation while ensuring necessary ensemble N-representability conditions \cite{Piris2018a}. For a long time, pure and ensemble universal functionals were believed to be equivalent within the domain of pure-state N-representable 1RDMs \cite{Nguyen-Dang1985}. However, more recent studies \cite{Schilling2018} revealed that the ensemble functional is actually the lower convex envelope of the pure functional, highlighting a fundamental distinction between the two. Fortunately, their equivalence was later confirmed on the set of v-representable 1RDMs \cite{Gritsenko2019}, ensuring that for ground-state electronic systems under an external potential $v(\bf r)$, pure and ensemble functionals are indistinguishable. As a result, we can impose only ensemble constraints during the energy minimization process, simplifying the enforcement of N-representability conditions.

PNOFs leverage the exact functional of the two-particle reduced density matrix ($D$) \cite{Husimi1940}, requiring only the reconstruction $D[n_{i},n_{j},n_{k},n_{l}]$ from the occupation numbers, leading to the following general form:

\begin{equation}
  E\left[\left\{ n_{i},\phi_{i}\right\} \right] = \sum\limits _{i}n_{i}H_{ii}+\sum\limits _{ijkl}D[n_{i},n_{j},n_{k},n_{l}]\left\langle kl|ij\right\rangle \label{ENOF}
\end{equation}
where $n_{i}$ denotes the occupation number of the natural spin orbital $\phi_{i}$. The term $H_{ii}$ represents the one-particle Hamiltonian matrix elements, including kinetic energy and potential energy contributions:

\begin{equation}
H_{ii} =\int d{\bf r} \phi _i ^*({\bf r})\left(-\frac{\nabla ^2 _r} {2} + v({\bf r}) \right) \phi _i ({\bf r})
\end{equation}
while $\left\langle kl|ij\right\rangle$ denotes the two-particle interaction matrix elements, given by:

\begin{equation}
    \langle kl|ij \rangle= \int \int d{\bf r}_1 d {\bf r}_2 \frac{\phi ^* _k ({\bf r}_1)\phi ^* _l ({\bf r}_2)\phi _i ({\bf r}_1)\phi _j({\bf r}_2)}{|{\bf r}_2 -{\bf r}_1|}
\end{equation}

The success of PNOFs lies in the fact that $D$ satisfies 2-positivity necessary ensemble N-representability conditions \cite{Mazziotti2012} and is reconstructed in terms of two-index auxiliary matrices, leading to JKL-only functionals of the following form: 
\begin{equation}    
    E = 2 \sum_p n_p H_{pp} + \sum_{pq} A[n_p,n_q] J_{qp} - \sum_{pq} B[n_p,n_q] K_{qp} - \sum_{pq} C[n_p,n_q] L_{qp} \label{PNOF}
\end{equation}
where $J_{qp}$, $K_{qp}$, and $L_{qp}$ represent the usual Coulomb, exchange, and exchange-time-inversion integrals \cite{Piris1999, Rodríguez-Mayorga2024}, respectively. Note that the indices here refer to natural spatial orbitals, and their occupancies $\left\{n_p\right\}$ determine the functions $A$, $B$, and $C$. The fundamental features of these functionals have previously been analyzed in several review articles \cite{Piris2013, Piris2014, Mitxelena2019}.

Notably, the PNOFs based on electron pairing have demonstrated significant potential \cite{Piris2018b} due to their ability to directly incorporate both dynamic and static correlation effects, allowing for a more accurate and computationally efficient description of strongly correlated systems. Within this framework, four functionals stand out: PNOF5 \cite{Piris2011, Piris2013a}, PNOF6 \cite{Piris2014a}, PNOF7 \cite{Piris2017,Mitxelena2018}, and the recently proposed global NOF (GNOF) \cite{Piris2021}. The electron-pairing methodology divides electron correlation into intrapair and interpair components. PNOF5 is an N-representable functional that accounts for intrapair electron correlation, while PNOF7 extends this description by including static interpair electron correlation effects. PNOF6, on the other hand, considers both intrapair and interpair correlations but fails to capture a significant portion of the correlation energy. GNOF addresses this limitation by incorporating dynamic interpair electron correlation, achieving a balance between static and dynamic correlation effects. Additional information on GNOF for singlet states, which are the only states considered in this work, can be found in the Appendix.

PNOF calculations can be performed using the DoNOF program \cite{Piris2021a}, an open-source software written in Fortran. This program is specifically designed to address the energy minimization problem for the ground-state energy (\ref{PNOF}), adhering to the orthonormality constraints of the natural orbitals and the electron-pairing restrictions on the occupation numbers. Importantly, these pairing constraints inherently satisfy the ensemble N-representability conditions of the one-particle reduced density matrix \cite{Coleman1963}. The optimization process is carried out in two interconnected stages, where the energy is independently optimized with respect to the occupation numbers and natural orbitals. Orbital optimization is achieved using either the iterative diagonalization method \cite{Piris2009} or the adaptive momentum technique recently introduced for this purpose \cite{Lew-Yee2025}. Additionally, specialized parameterizations have been developed \cite{Franco2024} to comply with the pairing constraints, enabling unconstrained optimization of the occupation numbers. These two stages are iteratively executed in an integrated process until convergence is reached for both the occupation numbers and the natural orbitals.

Despite the physical advantages of the electron-pairing scheme, its current implementation with orbitals of a space being contiguous imposes an artificial limitation, restricting flexibility in adjusting the number of orbitals coupled within subspaces to ensure electron pairing. This sorting makes it impossible to reuse previous calculations when the size of the electron-pairing scheme is increased. This work overcomes this drawback by introducing an alternative orbital sorting approach that alternates the sequence of orbital coupling. The proposed method retains the functional definitions while providing greater flexibility in defining subspaces. Notably, it enables the reuse of previously converged solutions for subsequent calculations, facilitating seamless adjustments to the number of coupled orbitals, whether increasing or decreasing them as required.

In this study, we compare the original coupling strategy, here referred to as ``continuous orbital sorting'', with the new ``alternating orbital sorting''. The proposed approach is validated through calculations on \ce{H2O}, \ce{H2O2} and \ce{NH3}, focusing on computational time and iteration count. We evaluated three scenarios: (i) a one-shot calculation, the ``old'' way,  where the PNOF computation is performed directly with the maximum number of coupled orbitals; (ii) an incremental approach, starting with a single coupled orbital and gradually increasing the number up to the maximum; and (iii) a two-step calculation, which begins with a single coupled orbital and then directly transitions to the maximum number of coupled orbitals. The latter two approaches, made possible by the new coupling strategy, demonstrate significant improvements in both efficiency and flexibility.

\section{Algorithm}

Typically, we consider $\mathrm{N_{I}}$ unpaired electrons, which determine the total spin of the system $S$. The remaining electrons, $\mathrm{N_{II}} = \mathrm{N-N_{I}}$, form electron pairs with opposite spins, resulting in a net spin of zero for the paired electrons. For the mixed state with the highest multiplicity, defined by $2S+1=\mathrm{N_{I}}+1$, the expected value of $\hat{S}_{z}$ is zero. As a result, the restricted spin formalism is applied, where $n_p^ {\alpha} =n_p^{\beta}=n_p$ \cite{Piris2019}.

In line with $\mathrm{N_{I}}$ and $\mathrm{N_{II}}$, the orbital space $\Omega$ is divided into two subspaces: $\Omega = \Omega_{\mathrm{I}} \oplus \Omega_{\mathrm{II}}$. $\Omega_{\mathrm{II}}$ is composed of $\mathrm{N_{II}}/2$ mutually disjoint subspaces $\Omega{}_{g}$. Each of which contains one strongly occupied orbital $\left|g\right\rangle $ with $g\leq\mathrm{N_{II}}/2$, and $\mathrm{N}_{g}$ weakly occupied orbitals $\left|p\right\rangle $ with $p>\mathrm{N_{II}}/2$, namely,
\begin{equation} 
\Omega_{g}=\left\{ \left|g\right\rangle ,\left|p_{1}\right\rangle ,\left|p_{2}\right\rangle ,...,\left|p_{\mathrm{N}_{g}}\right\rangle \right\} .\label{OmegaG}
\end{equation}

Taking into account the spin, the total occupancy for a given subspace $\Omega_{g}$ is 2, which is reflected in the following $\mathrm{N_{II}}/2$ pairing conditions:
\begin{equation}
\sum_{p\in\Omega_{g}}n_{p}=n_{g}+\sum_{i=1}^{\mathrm{N}_{g}}n_{p_{i}}=1,\quad g=1,2,...,\frac{\mathrm{N_{II}}}{2}.\label{sum1}
\end{equation}

In general, $\mathrm{N}_{g}$ can vary across subspaces as long as it adequately describes the electron pair. However, for simplicity, $\mathrm{N}_{g}$ is assumed to be the same for all subspaces $\Omega_{g}\in\Omega_{\mathrm{II}}$. The maximum allowable value of $\mathrm{N}_{g}$ is determined by the basis set employed in the calculations. From (\ref{sum1}), the following expression holds:
\begin{equation}
2\sum_{p\in\Omega_{\mathrm{II}}}n_{p}=2\sum_{g=1}^{\mathrm{N_{II}}/2}\left(n_{g}+\sum_{i=1}^{\mathrm{N}_{g}}n_{p_{i}}\right)=\mathrm{N_{II}}.\label{sumNpII}
\end{equation}

Similarly, $\Omega_{\mathrm{I}}$ is composed of $\mathrm{N_{I}}$ mutually disjoint subspaces $\Omega{}_{g}$. Unlike $\Omega_{\mathrm{II}}$, each subspace $\Omega_{g} \in \Omega_{\mathrm{I}}$ contains only one orbital $g$ with $n_{g}=1/2$. As expected, the orbitals in $\Omega_{\mathrm{I}}$ are not subject to the pairing condition, since each is singly occupied by an electron. However, the specific spin state of these electrons, whether $\alpha$ or $\beta$ remains indeterminate. Consequently, we have:
\begin{equation}
2\sum_{p\in\Omega_{\mathrm{I}}}n_{p}=2\sum_{g=\mathrm{N_{II}}/2+1}^{\mathrm{N_{II}}/2+\mathrm{N_{I}}}n_{g}=\mathrm{N_{I}}.\label{sumNpI}
\end{equation}
Taking into account Eqs. (\ref{sumNpII}) and (\ref{sumNpI}), the trace of the one-particle reduced density matrix is verified to be equal to the number of electrons: 
\begin{equation}
2\sum_{p\in\Omega}n_{p}=2\sum_{p\in\Omega_{\mathrm{II}}}n_{p}+2\sum_{p\in\Omega_{\mathrm{I}}}n_{p}=\mathrm{N_{II}}+\mathrm{N_{I}}=\mathrm{\mathrm{N}}.\label{norm}
\end{equation}

\begin{figure*}[htb]
    \centering
    \begin{subfigure}[b]{0.45\textwidth}
        \begin{tikzpicture}

    \draw[blue, dashed, thick] (0,10) rectangle ++(3,-2);
    \draw[blue,<->,very thick] (3.5,1) -- ++(1,0) -- ++(0,8) -- ++(-1,0) ;
    \node[blue,anchor=west] at (3.25,1.25) {$\Omega_1$};
    \draw [blue,very thick] (0.3,8.5) -- (2.75,8.5);
    \draw[->,very thick] (1,8.5) -- (1,8.6);
    \draw[<-,very thick] (2,8.5) -- (2,8.6);
    
    \draw [blue,very thick] (0.3,9.5) -- (2.75,9.5);
    \draw[->,very thick] (1,9.5) -- (1,9.6);
    \draw[<-,very thick] (2,9.5) -- (2,9.6);
    
    \draw [blue,very thick] (0.3,1) -- (2.75,1);   
    \draw[->,very thick] (1,1) -- (1,1.6);
    \draw[<-,very thick] (2,1) -- (2,1.6);

    \draw[Green, dashed, thick] (0,7.75) rectangle ++(3,-2); 
    \draw[Green,<->,very thick] (3.5,2) -- ++(0.75,0) -- ++(0,4.75) -- ++(-0.75,0) ;
    \node[Green,anchor=west] at (3.25,2.25) {$\Omega_2$};
    \draw [Green,very thick] (0.3,6.25) -- (2.75,6.25);
    \draw[->,very thick] (1,6.25) -- (1,6.4);
    \draw[<-,very thick] (2,6.25) -- (2,6.4);
    
    \draw [Green,very thick] (0.3,7.25) -- (2.75,7.25);
    \draw[->,very thick] (1,7.25) -- (1,7.4);
    \draw[<-,very thick] (2,7.25) -- (2,7.4);
    
    \draw [Green,very thick] (0.3,2) -- (2.75,2);
    \draw[->,very thick] (1,2) -- (1,2.5);
    \draw[<-,very thick] (2,2) -- (2,2.5);

    \draw[red, dashed, thick] (0,5.5) rectangle ++(3,-2); 
    \draw[red,<->,very thick] (3.5,3) -- ++(0.5,0) -- ++(0,1.5) -- ++(-0.5,0) ;
    \node[red,anchor=west] at (3.25,3.25) {$\Omega_3$};

    \draw [red,very thick] (0.3,4) -- (2.75,4);
    \draw[->,very thick] (1,4) -- (1,4.2);
    \draw[<-,very thick] (2,4) -- (2,4.2);
    
    \draw [red,very thick] (0.3,5) -- (2.75,5);
    \draw[->,very thick] (1,5) -- (1,5.2);
    \draw[<-,very thick] (2,5) -- (2,5.2);
    
    \draw [red,very thick] (0.3,3) -- (2.75,3);
    \draw[->,very thick] (1,3) -- (1,3.4);
    \draw[<-,very thick] (2,3) -- (2,3.4);

    \node[align=center] at (-2,8) { \\ };
    \node[align=center] at (-2,2) { \\ };
\end{tikzpicture}
        \caption{Continious Orbital Sorting (Old)}
        \label{fig:continuous-sorting}
    \end{subfigure}
    \hfill
    \begin{subfigure}[b]{0.5\textwidth}
        \begin{tikzpicture}

    \draw[dashed, thick] (0,10) rectangle ++(3,-3);
    \draw[dashed, thick] (0,6.75) rectangle ++(3,-3); 
    
    \draw[blue,<->,very thick] (3.5,1) -- ++(1,0) -- ++(0,5.25) -- ++(-1,0) ;
    \draw[blue,->,very thick](3.5,1) -- ++(1,0) -- ++(0,8.5) -- ++(-1,0);
    \node[blue,anchor=west] at (3.25,1.25) {$\Omega_1$};
    \draw [blue,very thick] (0.3,6.25) -- (2.75,6.25);
    \draw[->,very thick] (1,6.25) -- (1,6.35);
    \draw[<-,very thick] (2,6.25) -- (2,6.35);
    
    \draw [blue,very thick] (0.3,9.5) -- (2.75,9.5);
    \draw[->,very thick] (1,9.5) -- (1,9.6);
    \draw[<-,very thick] (2,9.5) -- (2,9.6);
    
    \draw [blue,very thick] (0.3,1) -- (2.75,1);   
    \draw[->,very thick] (1,1) -- (1,1.6);
    \draw[<-,very thick] (2,1) -- (2,1.6);

    \draw[Green,<->,very thick] (3.5,2) -- ++(0.75,0) -- ++(0,3) -- ++(-0.75,0) ;
    \draw[Green,->,very thick](3.5,2) -- ++(0.75,0) -- ++(0,6.25) -- ++(-0.75,0);
    \node[Green,anchor=west] at (3.25,2.25) {$\Omega_2$};
    
    \draw [Green,very thick] (0.3,5) -- (2.75,5);
    \draw[->,very thick] (1,5) -- (1,5.15);
    \draw[<-,very thick] (2,5) -- (2,5.15);
    
    \draw [Green,very thick] (0.3,8.25) -- (2.75,8.25);
    \draw[->,very thick] (1,8.25) -- (1,8.4);
    \draw[<-,very thick] (2,8.25) -- (2,8.4);
    
    \draw [Green,very thick] (0.3,2) -- (2.75,2);
    \draw[->,very thick] (1,2) -- (1,2.5);
    \draw[<-,very thick] (2,2) -- (2,2.5);

    \draw[red,<->,very thick](3.5,3) -- ++(0.5,0) -- ++(0,1) -- ++(-0.5,0) ;
    \draw[red,->,very thick](3.5,3) -- ++(0.5,0) -- ++(0,4.25) -- ++(-0.5,0);

    \node[red,anchor=west] at (3.25,3.25) {$\Omega_3$};

    \draw [red,very thick] (0.3,4) -- (2.75,4);
    \draw[->,very thick] (1,4) -- (1,4.2);
    \draw[<-,very thick] (2,4) -- (2,4.2);
    
    \draw [red,very thick]  (0.3,7.25) -- (2.75,7.25);
    \draw[->,very thick] (1,7.25) -- (1,7.45);
    \draw[<-,very thick] (2,7.25) -- (2,7.45);
    
    \draw [red,very thick] (0.3,3) -- (2.75,3);
    \draw[->,very thick] (1,3) -- (1,3.4);
    \draw[<-,very thick] (2,3) -- (2,3.4);

    \node[align=center,rotate=90] at (-1,7) {\small{Weakly Double} \\ \small{Occupied Orbitals}};
    \node[align=center,rotate=90] at (-1,2.5) {\small{Strongly Double} \\ \small{Occupied Orbitals}};

    \node[align=center] at (6,8) {\\ };
    \node[align=center] at (6,2) { \\ };
\end{tikzpicture}
        \caption{Alternating Orbital Sorting (New)}
        \label{fig:alternating-sorting}
    \end{subfigure}
    \caption{Graphical depiction of continuous orbital sorting and the alternating orbital sorting. In this example, $\mathrm{N_{I}} = 0$ ($S = 0$), whereas six electrons ($\mathrm{N_{II}} = 6$) distributed in three subspaces ${\Omega_{1}, \Omega_{2}, \Omega_{3}}$ make up the subspace $\Omega_{\mathrm{II}}$. Note that $\mathrm{N}_{g} = 2$. The arrows represent the occupation numbers for alpha ($\uparrow$) and beta ($\downarrow$) spins in each orbital}
    \label{fig:sorting}
\end{figure*}
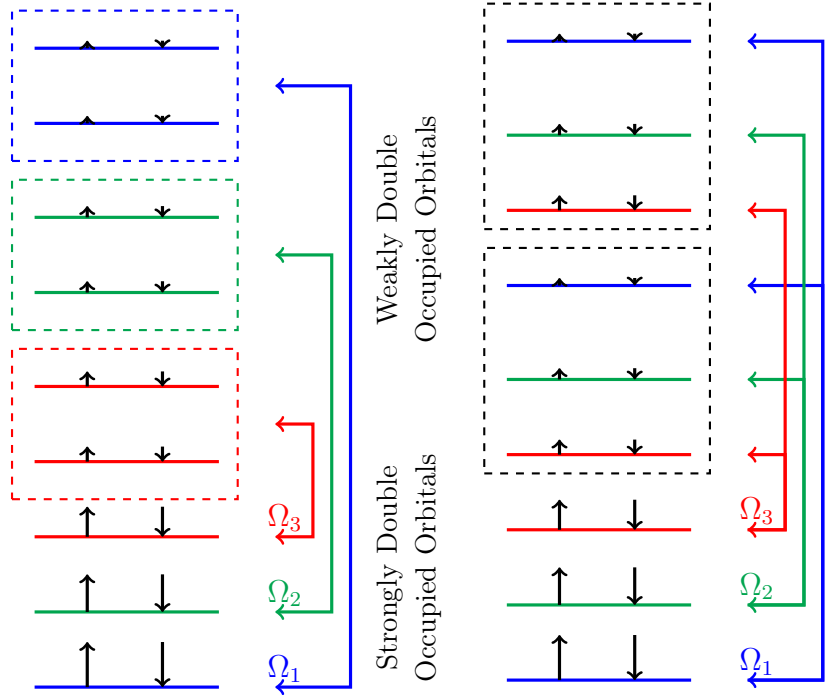

From this point forward, we will focus on the subspace $\Omega_{\mathrm{II}}$, as it is the one affected by the pairing conditions, whose implementation will be modified. Figure \ref{fig:sorting} provides an illustrative example. In this case, $\mathrm{N_{I}}=0$ ($S = 0$), indicating the absence of singly occupied orbitals, while six electrons ($\mathrm{N_{II}}=6$) are distributed across three subspaces, ${\Omega_{1}, \Omega_{2}, \Omega_{3}}$, which collectively constitute $\Omega_{\mathrm{II}}$. In this example, $\mathrm{N}_{g} = 2$ corresponds to the weakly occupied orbitals, which are depicted within the dashed-line boxes and are paired within each subspace $\Omega_{g}$ to the strongly occupied orbitals shown in the lower lines.

It is important to recall that the orbitals belonging to each subspace $\Omega_g$ vary throughout the optimization process until the most favorable orbital interactions are found. As a result, they do not remain fixed but adapt to the specific problem at hand, regardless of the starting orbitals used in the calculation. This means that the order in which the weakly occupied orbitals are selected to form the subspaces $\Omega_g$ does not affect the physical nature of the problem. However, it can influence the number of iterations and consequently the time required to obtain the optimal orbitals.

The simplest way to construct a subspace $\Omega_g$ is to select the $\mathrm{N_g}$ weakly occupied orbitals coupled to a strongly occupied orbital in a continuous manner, as illustrated in the dashed-line boxes in Figure~\ref{fig:continuous-sorting}. However, this continuous ordering introduces an artificial barrier when changing $\mathrm{N_g}$ for a new calculation. For example, in Figure~\ref{fig:continuous-sorting}, where $\mathrm{N_g}=2$, restarting the calculation with $\mathrm{N_g}=1$ using the previously converged orbitals from the $\mathrm{N_g}=2$ calculation as starting orbitals causes the second orbital in $\Omega_3$ to shift into $\Omega_2$, discarding the previously converged orbital in $\Omega_2$. Comparable situations will occur when, instead of decreasing $\mathrm{N_g}$ for a new calculation, the number of coupled orbitals is increased. Consequently, several additional iterations will be needed to regain convergence.

This issue can be addressed by using an alternating orbital sorting method. As illustrated in Figure~\ref{fig:alternating-sorting}, we assign the first weakly occupied orbital to $\Omega_3$, the next one to $\Omega_2$, and the following one to $\Omega_1$. This process is repeated $\mathrm{N_g}$ times. Since each dashed-box in the Figure~\ref{fig:alternating-sorting} contains an orbital from each subspace, adjusting the number of coupled orbitals in a subspace (either fewer or more) corresponds to reducing or increasing the number of dashed-boxes, respectively. This approach allows us to reuse the previously converged orbitals, which ultimately enhances the efficiency of the calculation.

\section{Results}

In this section, we present the results obtained using GNOF \cite{Piris2021}, our latest functional designed to achieve a balanced treatment of static and dynamic correlation effects. Orbital optimization was performed with the recently implemented adaptive momentum technique \cite{Lew-Yee2025}, while the unconstrained optimization of occupation numbers was performed using the softmax parametrization \cite{Franco2024}, ensuring compliance with the pairing constraints. All calculations were performed using experimental geometry and correlation-consistent cc-pVTZ basis sets developed by Dunning and coworkers \cite{Dunning1989}. A parallel version of the DoNOF software was used on 0na AMD Ryzen 5 4600H CPU.

\begin{figure}[htbp]
  \centering
  \begin{tikzpicture}[scale=1.2,font=\scriptsize]
    \begin{axis}[
      name=first plot,
      xticklabels={},
      xtick=\empty,
      ylabel={Time (s)},
      height=0.2\textwidth,
      width=0.62\textwidth,
      y tick label style={
          /pgf/number format/.cd,
          fixed,
          fixed zerofill,
          precision=3,
          /tikz/.cd
      },
      scaled y ticks=false,
      ylabel shift = 10 pt,
    ]
  \addplot[red, mark=*,smooth] table {
   1        0.007
   2        0.007
   3        0.007
   4        0.009
   5        0.010
   6        0.011
   7        0.013
   8        0.014
   9        0.014
  10        0.017
  11        0.018
  12        0.019
    };
    \end{axis}

    \begin{axis}[
        xlabel={$\mathrm{N_g}$},
        ylabel={Energy ($E_\text{h}$)},
        xmin=1,
        ymax=-76.2,
        y tick label style={
          /pgf/number format/.cd,
          fixed,
          fixed zerofill,
          precision=2,
          /tikz/.cd
      },
        at=(first plot.south),
        anchor=north,
        yshift=-5mm,
    ]
    \addplot table [x=ncwo, y=incremental, smooth] {h2o-energy.dat};

    \node[anchor=west] (source1) at (2.0,-76.24){Perfect Pairing};
    \node (destination1) at (1,-76.23){};
    \draw[->](source1)--(destination1);

    \node[anchor=west] (source2) at (10.6,-76.33){$\mathrm{N^{max}_g}$};
    \node (destination2) at (12,-76.347){};
    \draw[->](source2)--(destination2);

    \draw [decorate, decoration={brace,amplitude=20pt,raise=2pt}] (2,-76.290) -- (12,-76.290) node [midway, anchor=south west, yshift=4mm, xshift=-22.8mm, outer sep=10pt, rotate=0] {Extended Pairing};
    
    \end{axis}
\end{tikzpicture}
  \caption{\ce{H2O} calculated using GNOF/cc-pVTZ as a function of the subspace. The one-shot scheme was employed, that is, each point has been computed independently. Top panel shows the time per orbital iteration, while bottom panel shows the energy for each subspace size.}
  \label{fig:energy_by_subspace}
\end{figure}
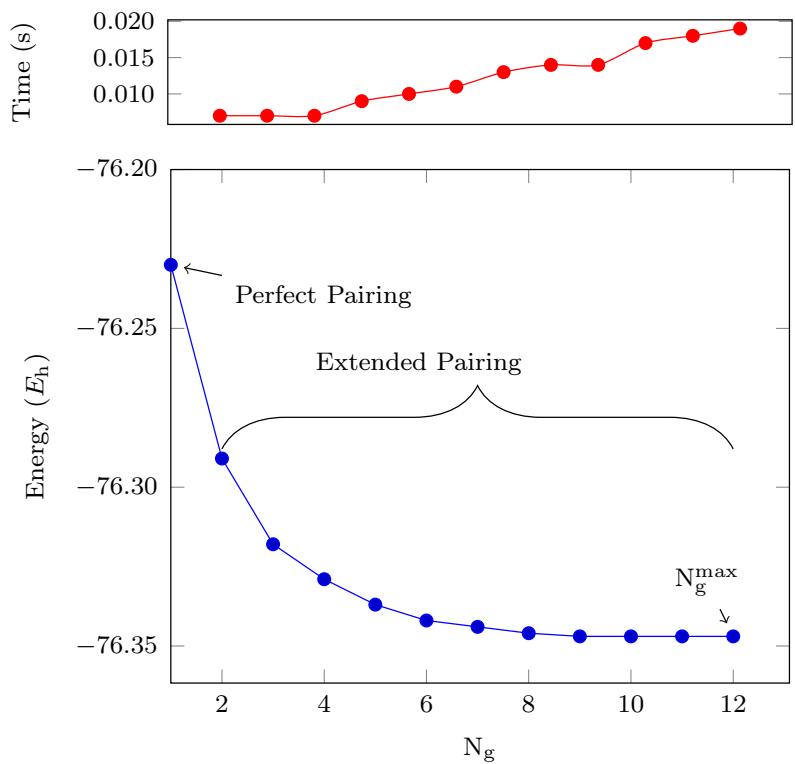

The water molecule is one of the most studied systems in quantum chemistry, making it an ideal candidate for examining how subspace size influences electronic structure calculations. It also provides a suitable benchmark to assess the performance of the new alternating orbital sorting implemented in the DoNOF program \cite{Piris2021a} compared to continuous orbital sorting. As a benchmark system, water enables a systematic investigation of how subspace size affects the convergence behavior and accuracy of NOF calculations, as well as how increasing the number of weakly occupied orbitals impacts the recovery of electron correlation. To illustrate this, we calculated the energy of water for different values of $\mathrm{N_g}$, recalling that $\mathrm{N_g}$ represents the number of weakly occupied orbitals coupled with strongly occupied ones in each subspace. The effect of subspace size on the total electronic energy is depicted in Figure~\ref{fig:energy_by_subspace}, where each point corresponds to an independent one-shot calculation, which means that the optimization was performed from scratch for each selected value of $\mathrm{N_g}$.

In the simplest case, where only one weakly occupied orbital is coupled with a strongly occupied orbital in each subspace ($\mathrm{N_g}=1$), the calculation predominantly captures static correlation effects. This scenario, known as perfect pairing, maximizes the fractional occupation of the weakly occupied orbital while ensuring strict electron pairing. In systems dominated by static correlation, such as dissociations or diradical species, this configuration provides an accurate description of the electronic structure, with weakly occupied orbitals reaching half-occupation. However, for systems where dynamic correlation is significant, perfect pairing alone is insufficient, as it fails to describe the finer energy contributions arising from electron interactions beyond strictly paired orbitals.

To recover these missing correlation effects, it is necessary to expand the subspace by incorporating additional weakly occupied orbitals ($\mathrm{N_g}>1$). This extension follows the extended pairing scheme, where more orbitals participate in the correlation process, enabling a more balanced description of both static and dynamic contributions. As shown in Figure~\ref{fig:energy_by_subspace}, increasing $\mathrm{N_g}$ systematically lowers the electronic energy, reflecting the gradual recovery of correlation energy. While larger subspaces improve accuracy, they also increase computational cost, as the number of variational parameters and required optimization steps grows significantly.

The results shown in Figure~\ref{fig:energy_by_subspace} underscore the fundamental role of the subspace size in capturing electron correlation effects within NOFT. Ideally, the highest accuracy is achieved when the maximum number of orbitals allowed by the basis set is included. For the \ce{H2O} molecule with the cc-pVTZ basis set, this corresponds to $\mathrm{N^{max}_g}=12$. The observed energy trend confirms that increasing $\mathrm{N_g}$ systematically improves the accuracy of the electronic structure description by incorporating more dynamic correlation. However, this improvement comes at the cost of increased computational effort, highlighting the need for efficient strategies to optimize orbital selection and convergence. Although the results presented here were obtained using the one-shot scheme, they establish a crucial baseline for assessing the impact of different orbital sorting strategies, which will be discussed below.

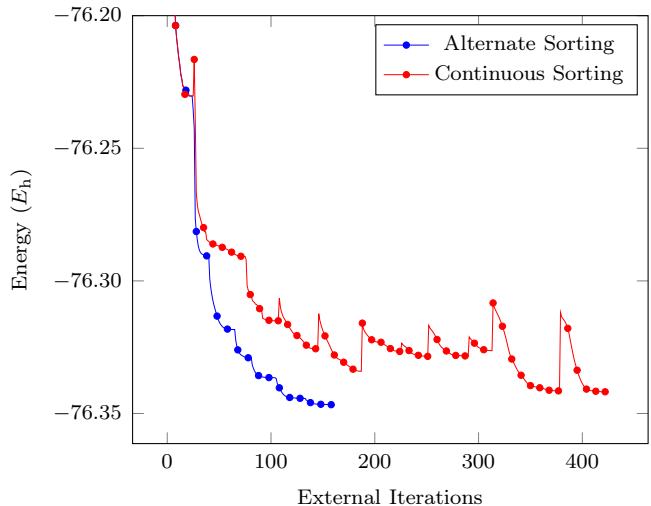
\begin{figure}[htbp]
    \centering
    \begin{tikzpicture}[font=\scriptsize]
  \begin{axis}[
    xlabel={External Iterations},
    ylabel={Energy ($E_\text{h}$)},
    ymax=-76.2,
    y tick label style={
        /pgf/number format/.cd,
        fixed,
        fixed zerofill,
        precision=2,
        /tikz/.cd
    },
  ]
  \addplot[blue, mark=*, line width=0.001pt, mark size=1.2pt, mark repeat=10] table {
   0  -70.92550
   1  -73.55497
   2  -74.80562
   3  -75.49185
   4  -75.88558
   5  -76.07463
   6  -76.15667
   7  -76.19199
   8  -76.20372
   9  -76.20839
  10  -76.21183
  11  -76.21563
  12  -76.21851
  13  -76.22164
  14  -76.22380
  15  -76.22637
  16  -76.22711
  17  -76.22761
  18  -76.22810
  19  -76.22842
  20  -76.22807
  21  -76.23028
  22  -76.23030
  23  -76.23030
  24  -76.23030
  25  -76.23563
  26  -76.24244
  27  -76.27651
  28  -76.28140
  29  -76.28471
  30  -76.28703
  31  -76.28848
  32  -76.28931
  33  -76.28978
  34  -76.29005
  35  -76.29020
  36  -76.29032
  37  -76.29046
  38  -76.29059
  39  -76.29068
  40  -76.29068
  41  -76.29938
  42  -76.30280
  43  -76.30530
  44  -76.30741
  45  -76.30914
  46  -76.31060
  47  -76.31196
  48  -76.31329
  49  -76.31447
  50  -76.31543
  51  -76.31617
  52  -76.31674
  53  -76.31720
  54  -76.31756
  55  -76.31780
  56  -76.31798
  57  -76.31810
  58  -76.31819
  59  -76.31825
  60  -76.31829
  61  -76.31831
  62  -76.31833
  63  -76.31834
  64  -76.31835
  65  -76.31835
  66  -76.32234
  67  -76.32457
  68  -76.32603
  69  -76.32691
  70  -76.32751
  71  -76.32789
  72  -76.32816
  73  -76.32840
  74  -76.32860
  75  -76.32875
  76  -76.32887
  77  -76.32895
  78  -76.32901
  79  -76.32904
  80  -76.32907
  81  -76.32907
  82  -76.33185
  83  -76.33294
  84  -76.33389
  85  -76.33459
  86  -76.33509
  87  -76.33546
  88  -76.33572
  89  -76.33592
  90  -76.33608
  91  -76.33618
  92  -76.33627
  93  -76.33633
  94  -76.33638
  95  -76.33641
  96  -76.33643
  97  -76.33645
  98  -76.33647
  99  -76.33649
 100  -76.33650
 101  -76.33651
 102  -76.33652
 103  -76.33652
 104  -76.33655
 105  -76.33685
 106  -76.33863
 107  -76.33950
 108  -76.34033
 109  -76.34109
 110  -76.34178
 111  -76.34238
 112  -76.34286
 113  -76.34323
 114  -76.34350
 115  -76.34370
 116  -76.34383
 117  -76.34392
 118  -76.34398
 119  -76.34403
 120  -76.34407
 121  -76.34411
 122  -76.34414
 123  -76.34417
 124  -76.34420
 125  -76.34423
 126  -76.34425
 127  -76.34428
 128  -76.34429
 129  -76.34431
 130  -76.34432
 131  -76.34432
 132  -76.34433
 133  -76.34435
 134  -76.34499
 135  -76.34521
 136  -76.34547
 137  -76.34571
 138  -76.34592
 139  -76.34609
 140  -76.34622
 141  -76.34631
 142  -76.34639
 143  -76.34644
 144  -76.34648
 145  -76.34650
 146  -76.34652
 147  -76.34653
 148  -76.34655
 149  -76.34657
 150  -76.34659
 151  -76.34660
 152  -76.34661
 153  -76.34662
 154  -76.34663
 155  -76.34663
 156  -76.34665
 157  -76.34667
 158  -76.34670
    };
    \addlegendentry{Alternate Sorting}

  \addplot[red, mark=*, line width=0.001pt, mark size=1.2pt, mark repeat=9] table {
   0  -70.92521
   1  -73.55479
   2  -74.80538
   3  -75.49104
   4  -75.88404
   5  -76.07369
   6  -76.15623
   7  -76.19027
   8  -76.20372
   9  -76.20839
  10  -76.21204
  11  -76.21555
  12  -76.21799
  13  -76.22243
  14  -76.22390
  15  -76.22663
  16  -76.22674
  17  -76.22967
  18  -76.22985
  19  -76.22998
  20  -76.23008
  21  -76.23016
  22  -76.23024
  23  -76.23028
  24  -76.23030
  25  -76.23030
  26  -76.21651
  27  -76.23948
  28  -76.26683
  29  -76.27134
  30  -76.27384
  31  -76.27506
  32  -76.27668
  33  -76.27805
  34  -76.27936
  35  -76.27991
  36  -76.28147
  37  -76.28147
  38  -76.28444
  39  -76.28482
  40  -76.28516
  41  -76.28546
  42  -76.28571
  43  -76.28592
  44  -76.28610
  45  -76.28626
  46  -76.28640
  47  -76.28653
  48  -76.28665
  49  -76.28677
  50  -76.28690
  51  -76.28704
  52  -76.28719
  53  -76.28735
  54  -76.28751
  55  -76.28768
  56  -76.28785
  57  -76.28802
  58  -76.28821
  59  -76.28842
  60  -76.28864
  61  -76.28890
  62  -76.28917
  63  -76.28944
  64  -76.28971
  65  -76.28997
  66  -76.29019
  67  -76.29037
  68  -76.29051
  69  -76.29061
  70  -76.29068
  71  -76.29073
  72  -76.29076
  73  -76.29078
  74  -76.29079
  75  -76.29079
  76  -76.29195
  77  -76.30182
  78  -76.30347
  79  -76.30429
  80  -76.30517
  81  -76.30566
  82  -76.30730
  83  -76.30787
  84  -76.30816
  85  -76.30878
  86  -76.30889
  87  -76.30942
  88  -76.31033
  89  -76.31048
  90  -76.31136
  91  -76.31121
  92  -76.31413
  93  -76.31431
  94  -76.31447
  95  -76.31460
  96  -76.31470
  97  -76.31479
  98  -76.31487
  99  -76.31493
 100  -76.31498
 101  -76.31501
 102  -76.31503
 103  -76.31505
 104  -76.31506
 105  -76.31506
 106  -76.31506
 107  -76.31506
 108  -76.30648
 109  -76.30971
 110  -76.31166
 111  -76.31284
 112  -76.31370
 113  -76.31443
 114  -76.31510
 115  -76.31577
 116  -76.31645
 117  -76.31712
 118  -76.31777
 119  -76.31839
 120  -76.31897
 121  -76.31948
 122  -76.31989
 123  -76.32020
 124  -76.32042
 125  -76.32063
 126  -76.32088
 127  -76.32118
 128  -76.32152
 129  -76.32190
 130  -76.32233
 131  -76.32280
 132  -76.32329
 133  -76.32378
 134  -76.32425
 135  -76.32467
 136  -76.32501
 137  -76.32526
 138  -76.32540
 139  -76.32548
 140  -76.32553
 141  -76.32557
 142  -76.32559
 143  -76.32561
 144  -76.32562
 145  -76.32562
 146  -76.31237
 147  -76.31477
 148  -76.31632
 149  -76.31760
 150  -76.31876
 151  -76.31981
 152  -76.32076
 153  -76.32163
 154  -76.32242
 155  -76.32320
 156  -76.32402
 157  -76.32488
 158  -76.32574
 159  -76.32655
 160  -76.32730
 161  -76.32796
 162  -76.32852
 163  -76.32896
 164  -76.32927
 165  -76.32949
 166  -76.32968
 167  -76.32988
 168  -76.33012
 169  -76.33039
 170  -76.33069
 171  -76.33099
 172  -76.33130
 173  -76.33160
 174  -76.33190
 175  -76.33219
 176  -76.33249
 177  -76.33278
 178  -76.33306
 179  -76.33331
 180  -76.33353
 181  -76.33371
 182  -76.33384
 183  -76.33394
 184  -76.33401
 185  -76.33406
 186  -76.33409
 187  -76.33409
 188  -76.31595
 189  -76.31790
 190  -76.31888
 191  -76.31954
 192  -76.32011
 193  -76.32060
 194  -76.32107
 195  -76.32151
 196  -76.32190
 197  -76.32222
 198  -76.32248
 199  -76.32268
 200  -76.32284
 201  -76.32297
 202  -76.32306
 203  -76.32311
 204  -76.32314
 205  -76.32315
 206  -76.32315
 207  -76.32315
 208  -76.32341
 209  -76.32396
 210  -76.32426
 211  -76.32450
 212  -76.32474
 213  -76.32500
 214  -76.32527
 215  -76.32553
 216  -76.32578
 217  -76.32602
 218  -76.32622
 219  -76.32637
 220  -76.32647
 221  -76.32654
 222  -76.32658
 223  -76.32661
 224  -76.32662
 225  -76.32662
 226  -76.32351
 227  -76.32408
 228  -76.32441
 229  -76.32471
 230  -76.32503
 231  -76.32540
 232  -76.32582
 233  -76.32629
 234  -76.32674
 235  -76.32712
 236  -76.32740
 237  -76.32759
 238  -76.32771
 239  -76.32780
 240  -76.32789
 241  -76.32798
 242  -76.32807
 243  -76.32815
 244  -76.32821
 245  -76.32827
 246  -76.32833
 247  -76.32838
 248  -76.32842
 249  -76.32845
 250  -76.32847
 251  -76.32847
 252  -76.31687
 253  -76.31769
 254  -76.31822
 255  -76.31872
 256  -76.31927
 257  -76.31987
 258  -76.32054
 259  -76.32128
 260  -76.32212
 261  -76.32298
 262  -76.32376
 263  -76.32440
 264  -76.32490
 265  -76.32530
 266  -76.32563
 267  -76.32590
 268  -76.32616
 269  -76.32644
 270  -76.32672
 271  -76.32701
 272  -76.32729
 273  -76.32754
 274  -76.32776
 275  -76.32793
 276  -76.32804
 277  -76.32809
 278  -76.32812
 279  -76.32815
 280  -76.32816
 281  -76.32818
 282  -76.32820
 283  -76.32822
 284  -76.32824
 285  -76.32826
 286  -76.32829
 287  -76.32831
 288  -76.32834
 289  -76.32836
 290  -76.32836
 291  -76.32115
 292  -76.32180
 293  -76.32218
 294  -76.32259
 295  -76.32303
 296  -76.32350
 297  -76.32397
 298  -76.32440
 299  -76.32477
 300  -76.32510
 301  -76.32533
 302  -76.32552
 303  -76.32568
 304  -76.32583
 305  -76.32597
 306  -76.32607
 307  -76.32615
 308  -76.32620
 309  -76.32624
 310  -76.32627
 311  -76.32630
 312  -76.32632
 313  -76.32632
 314  -76.30834
 315  -76.30978
 316  -76.31075
 317  -76.31156
 318  -76.31233
 319  -76.31314
 320  -76.31402
 321  -76.31497
 322  -76.31601
 323  -76.31713
 324  -76.31833
 325  -76.31961
 326  -76.32107
 327  -76.32268
 328  -76.32436
 329  -76.32599
 330  -76.32745
 331  -76.32859
 332  -76.32948
 333  -76.33025
 334  -76.33100
 335  -76.33178
 336  -76.33258
 337  -76.33335
 338  -76.33404
 339  -76.33462
 340  -76.33513
 341  -76.33559
 342  -76.33608
 343  -76.33659
 344  -76.33711
 345  -76.33763
 346  -76.33813
 347  -76.33859
 348  -76.33898
 349  -76.33929
 350  -76.33950
 351  -76.33964
 352  -76.33974
 353  -76.33983
 354  -76.33992
 355  -76.34001
 356  -76.34009
 357  -76.34017
 358  -76.34025
 359  -76.34032
 360  -76.34039
 361  -76.34046
 362  -76.34056
 363  -76.34068
 364  -76.34082
 365  -76.34095
 366  -76.34107
 367  -76.34117
 368  -76.34125
 369  -76.34132
 370  -76.34136
 371  -76.34139
 372  -76.34141
 373  -76.34143
 374  -76.34145
 375  -76.34147
 376  -76.34149
 377  -76.34150
 378  -76.34150
 379  -76.31190
 380  -76.31346
 381  -76.31409
 382  -76.31460
 383  -76.31520
 384  -76.31590
 385  -76.31677
 386  -76.31792
 387  -76.31953
 388  -76.32148
 389  -76.32358
 390  -76.32562
 391  -76.32756
 392  -76.32936
 393  -76.33096
 394  -76.33238
 395  -76.33369
 396  -76.33492
 397  -76.33605
 398  -76.33706
 399  -76.33795
 400  -76.33872
 401  -76.33939
 402  -76.33997
 403  -76.34045
 404  -76.34082
 405  -76.34107
 406  -76.34122
 407  -76.34131
 408  -76.34138
 409  -76.34145
 410  -76.34150
 411  -76.34155
 412  -76.34160
 413  -76.34163
 414  -76.34167
 415  -76.34171
 416  -76.34174
 417  -76.34177
 418  -76.34180
 419  -76.34182
 420  -76.34183
 421  -76.34184
 422  -76.34184
  };
    \addlegendentry{Continuous Sorting}

  \end{axis}
\end{tikzpicture}
    \caption{Energy variations in the water molecule during the incremental increase of $\mathrm{N_g}$, starting with a single weakly occupied orbital ($\mathrm{N_g}=1$) and reaching the maximum value ($\mathrm{N_g}=12$) allowed by the cc-pVTZ basis set.}
    \label{fig:incremental-old-vs-new}
\end{figure}

PNOFs provide a convenient polynomial scaling for managing large subspaces compared to traditional active-space wavefunction methods. Still, perfect pairing is a notable case, as it is computationally efficient due to the small subspace size, which allows for fast iterations and accelerates convergence by limiting the degrees of freedom. Therefore, starting with a perfect pairing calculation as an initial guess can be advantageous, followed by expanding the subspace size to expedite the computation. This approach, now labeled the incremental approach, is illustrated in Figure~\ref{fig:incremental-old-vs-new}. The figure compares two orbital sorting schemes: the previously employed continuous orbital sorting, shown in red, and the newly proposed alternating orbital sorting, represented in blue.

In the continuous orbital sorting scheme, the weakly occupied orbitals are grouped sequentially within subspaces. As a consequence, increasing the subspace size for a new calculation requires modifying the existing subspace structure, redistributing orbitals, and disrupting the previously converged solution. This results in a temporary increase in energy whenever an orbital is added, as the convergence process must restart to adjust to the new configuration. This behavior is evident in the red curve of Figure~\ref{fig:incremental-old-vs-new}, where the energy fluctuates and does not decrease monotonically as the number of coupled orbitals increases.

In contrast, the alternating orbital sorting method circumvents this issue by distributing weakly occupied orbitals across different subspaces in a staggered fashion. This ensures that newly added orbitals do not interfere with the previously established convergence, allowing for a seamless extension of the subspace. The effect of this strategy is reflected in the blue curve of Figure~\ref{fig:incremental-old-vs-new}, where the energy exhibits a smooth and continuous decrease as more orbitals are incorporated, leading to a more stable and efficient convergence process. Notably, this behavior demonstrates that the new sorting method effectively preserves the information obtained in previous calculations, reducing the computational overhead associated with recalculations.

The improvements achieved with the alternating orbital sorting method translate into a direct gain in computational efficiency. By enabling a smooth and monotonic energy descent, this approach significantly reduces the number of iterations required to reach convergence, compared to continuous orbital sorting. Consequently, the method facilitates a more flexible and scalable strategy for handling large subspaces while maintaining numerical stability. The findings presented in Figure~\ref{fig:incremental-old-vs-new} highlight the clear advantages of the new sorting approach, making it particularly valuable for studies involving strongly correlated systems that require efficient orbital optimization.

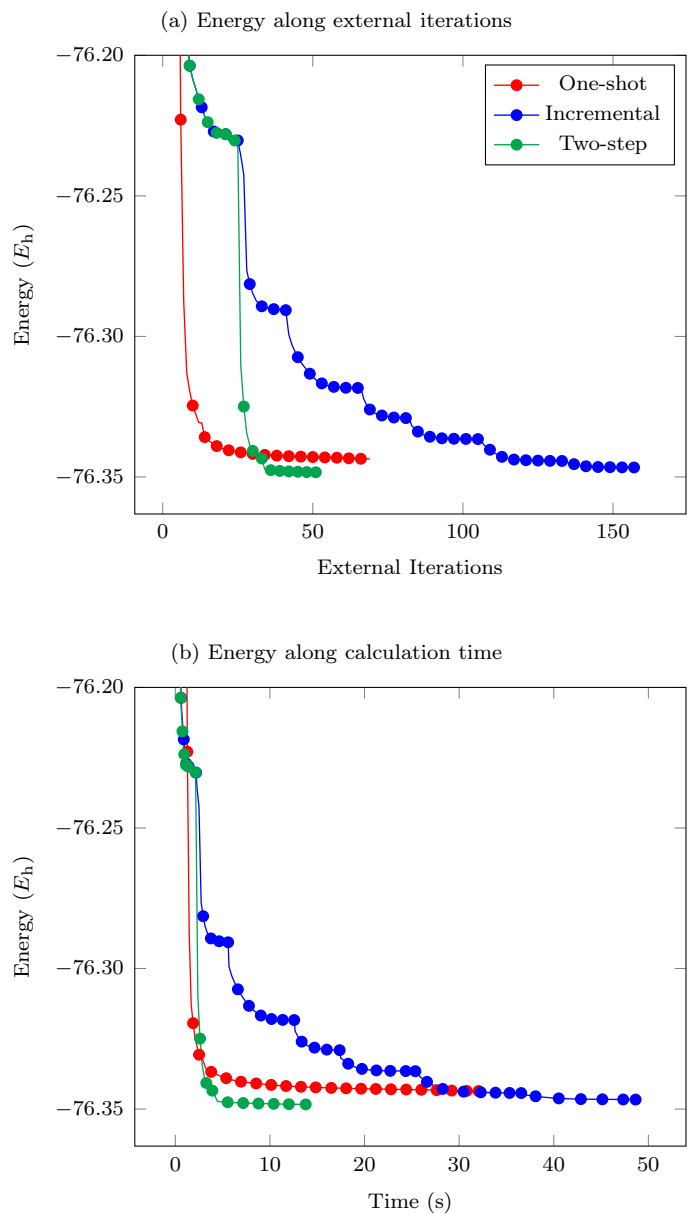
\begin{figure}[htbp]
  \centering
  \begin{subfigure}{0.65\textwidth}
    \caption{Energy along external iterations}
    \begin{tikzpicture}[font=\scriptsize]
  \begin{axis}[
    xlabel={External Iterations},
    ylabel={Energy ($E_\text{h}$)},
    ymax=-76.2,
    y tick label style={
        /pgf/number format/.cd,
        fixed,
        fixed zerofill,
        precision=2,
        /tikz/.cd
    },
    width=\textwidth,
  ]
  \addplot[red, mark=*, line width=0.5pt, mark size=2pt, mark repeat=4] table {
   1  -68.82353
   2  -72.02530
   3  -74.65078
   4  -75.75902
   5  -76.07990
   6  -76.22289
   7  -76.28762
   8  -76.31289
   9  -76.31945
  10  -76.32462
  11  -76.32767
  12  -76.33068
  13  -76.33079
  14  -76.33585
  15  -76.33679
  16  -76.33765
  17  -76.33841
  18  -76.33905
  19  -76.33957
  20  -76.33997
  21  -76.34028
  22  -76.34053
  23  -76.34073
  24  -76.34091
  25  -76.34109
  26  -76.34127
  27  -76.34143
  28  -76.34158
  29  -76.34171
  30  -76.34182
  31  -76.34192
  32  -76.34201
  33  -76.34209
  34  -76.34217
  35  -76.34224
  36  -76.34231
  37  -76.34238
  38  -76.34244
  39  -76.34250
  40  -76.34255
  41  -76.34259
  42  -76.34264
  43  -76.34268
  44  -76.34272
  45  -76.34277
  46  -76.34280
  47  -76.34284
  48  -76.34287
  49  -76.34291
  50  -76.34294
  51  -76.34298
  52  -76.34301
  53  -76.34305
  54  -76.34307
  55  -76.34310
  56  -76.34312
  57  -76.34315
  58  -76.34319
  59  -76.34324
  60  -76.34329
  61  -76.34334
  62  -76.34339
  63  -76.34344
  64  -76.34348
  65  -76.34352
  66  -76.34355
  67  -76.34358
  68  -76.34361
  69  -76.34361
    };
    \addlegendentry{One-shot}

  \addplot[blue, mark=*, line width=0.5pt, mark size=2pt, mark repeat=4] table {
   1  -70.92550
   2  -73.55497
   3  -74.80562
   4  -75.49185
   5  -75.88558
   6  -76.07463
   7  -76.15667
   8  -76.19199
   9  -76.20372
  10  -76.20839
  11  -76.21183
  12  -76.21563
  13  -76.21851
  14  -76.22164
  15  -76.22380
  16  -76.22637
  17  -76.22711
  18  -76.22761
  19  -76.22810
  20  -76.22842
  21  -76.22807
  22  -76.23028
  23  -76.23030
  24  -76.23030
  25  -76.23030
  26  -76.23563
  27  -76.24244
  28  -76.27651
  29  -76.28140
  30  -76.28471
  31  -76.28703
  32  -76.28848
  33  -76.28931
  34  -76.28978
  35  -76.29005
  36  -76.29020
  37  -76.29032
  38  -76.29046
  39  -76.29059
  40  -76.29068
  41  -76.29068
  42  -76.29938
  43  -76.30280
  44  -76.30530
  45  -76.30741
  46  -76.30914
  47  -76.31060
  48  -76.31196
  49  -76.31329
  50  -76.31447
  51  -76.31543
  52  -76.31617
  53  -76.31674
  54  -76.31720
  55  -76.31756
  56  -76.31780
  57  -76.31798
  58  -76.31810
  59  -76.31819
  60  -76.31825
  61  -76.31829
  62  -76.31831
  63  -76.31833
  64  -76.31834
  65  -76.31835
  66  -76.31835
  67  -76.32234
  68  -76.32457
  69  -76.32603
  70  -76.32691
  71  -76.32751
  72  -76.32789
  73  -76.32816
  74  -76.32840
  75  -76.32860
  76  -76.32875
  77  -76.32887
  78  -76.32895
  79  -76.32901
  80  -76.32904
  81  -76.32907
  82  -76.32907
  83  -76.33185
  84  -76.33294
  85  -76.33389
  86  -76.33459
  87  -76.33509
  88  -76.33546
  89  -76.33572
  90  -76.33592
  91  -76.33608
  92  -76.33618
  93  -76.33627
  94  -76.33633
  95  -76.33638
  96  -76.33641
  97  -76.33643
  98  -76.33645
  99  -76.33647
 100  -76.33649
 101  -76.33650
 102  -76.33651
 103  -76.33652
 104  -76.33652
 105  -76.33655
 106  -76.33685
 107  -76.33863
 108  -76.33950
 109  -76.34033
 110  -76.34109
 111  -76.34178
 112  -76.34238
 113  -76.34286
 114  -76.34323
 115  -76.34350
 116  -76.34370
 117  -76.34383
 118  -76.34392
 119  -76.34398
 120  -76.34403
 121  -76.34407
 122  -76.34411
 123  -76.34414
 124  -76.34417
 125  -76.34420
 126  -76.34423
 127  -76.34425
 128  -76.34428
 129  -76.34429
 130  -76.34431
 131  -76.34432
 132  -76.34432
 133  -76.34433
 134  -76.34435
 135  -76.34499
 136  -76.34521
 137  -76.34547
 138  -76.34571
 139  -76.34592
 140  -76.34609
 141  -76.34622
 142  -76.34631
 143  -76.34639
 144  -76.34644
 145  -76.34648
 146  -76.34650
 147  -76.34652
 148  -76.34653
 149  -76.34655
 150  -76.34657
 151  -76.34659
 152  -76.34660
 153  -76.34661
 154  -76.34662
 155  -76.34663
 156  -76.34663
 157  -76.34665
 158  -76.34667
 159  -76.34670
  };
    \addlegendentry{Incremental}

  \addplot[Green, mark=*, line width=0.5pt, mark size=2pt, mark repeat=3] table {
   1  -70.92550
   2  -73.55497
   3  -74.80562
   4  -75.49185
   5  -75.88558
   6  -76.07463
   7  -76.15667
   8  -76.19199
   9  -76.20372
  10  -76.20839
  11  -76.21183
  12  -76.21563
  13  -76.21851
  14  -76.22164
  15  -76.22380
  16  -76.22637
  17  -76.22711
  18  -76.22761
  19  -76.22810
  20  -76.22842
  21  -76.22807
  22  -76.23028
  23  -76.23030
  24  -76.23030
  25  -76.23030
  26  -76.31017
  27  -76.32496
  28  -76.33445
  29  -76.33880
  30  -76.34074
  31  -76.34253
  32  -76.34357
  33  -76.34347
  34  -76.34728
  35  -76.34746
  36  -76.34760
  37  -76.34772
  38  -76.34781
  39  -76.34788
  40  -76.34794
  41  -76.34800
  42  -76.34805
  43  -76.34809
  44  -76.34813
  45  -76.34818
  46  -76.34822
  47  -76.34825
  48  -76.34829
  49  -76.34831
  50  -76.34834
  51  -76.34836
  52  -76.34836
  };
    \addlegendentry{Two-step}
    
  \end{axis}
\end{tikzpicture}
  \end{subfigure}
  \hfill
  \begin{subfigure}{0.65\textwidth}
    \caption{Energy along calculation time}
    \begin{tikzpicture}[font=\scriptsize]
  \begin{axis}[
    xlabel={Time (s)},
    ylabel={Energy ($E_\text{h}$)},
    ymax=-76.2,
    y tick label style={
        /pgf/number format/.cd,
        fixed,
        fixed zerofill,
        precision=2,
        /tikz/.cd
    },
    width=\textwidth,
  ]
  \addplot[red, mark=*, line width=0.5pt, mark size=2pt, mark repeat=3] table {
  0.220  -68.82353
  0.440  -72.02530
  0.640  -74.65078
  0.850  -75.75902
  1.060  -76.07990
  1.270  -76.22289
  1.470  -76.28762
  1.680  -76.31289
  1.880  -76.31945
  2.090  -76.32462
  2.290  -76.32767
  2.510  -76.33068
  2.720  -76.33079
  3.260  -76.33585
  3.790  -76.33679
  4.320  -76.33765
  4.860  -76.33841
  5.380  -76.33905
  5.900  -76.33957
  6.430  -76.33997
  6.940  -76.34028
  7.490  -76.34053
  8.030  -76.34073
  8.550  -76.34091
  9.080  -76.34109
  9.610  -76.34127
 10.130  -76.34143
 10.650  -76.34158
 11.190  -76.34171
 11.710  -76.34182
 12.240  -76.34192
 12.760  -76.34201
 13.280  -76.34209
 13.800  -76.34217
 14.320  -76.34224
 14.840  -76.34231
 15.430  -76.34238
 15.950  -76.34244
 16.480  -76.34250
 17.010  -76.34255
 17.540  -76.34259
 18.060  -76.34264
 18.600  -76.34268
 19.120  -76.34272
 19.640  -76.34277
 20.170  -76.34280
 20.680  -76.34284
 21.200  -76.34287
 21.730  -76.34291
 22.260  -76.34294
 22.790  -76.34298
 23.320  -76.34301
 23.860  -76.34305
 24.410  -76.34307
 24.930  -76.34310
 25.450  -76.34312
 26.000  -76.34315
 26.550  -76.34319
 27.080  -76.34324
 27.620  -76.34329
 28.160  -76.34334
 28.680  -76.34339
 29.210  -76.34344
 29.740  -76.34348
 30.260  -76.34352
 30.800  -76.34355
 31.330  -76.34358
 32.040  -76.34361
 32.054  -76.34361
    };

  \addplot[blue, mark=*, line width=0.5pt, mark size=2pt, mark repeat=4] table {
  0.061  -70.92550
  0.123  -73.55497
  0.190  -74.80562
  0.252  -75.49185
  0.315  -75.88558
  0.381  -76.07463
  0.471  -76.15667
  0.538  -76.19199
  0.609  -76.20372
  0.680  -76.20839
  0.752  -76.21183
  0.821  -76.21563
  0.897  -76.21851
  0.964  -76.22164
  1.027  -76.22380
  1.088  -76.22637
  1.150  -76.22711
  1.213  -76.22761
  1.279  -76.22810
  1.341  -76.22842
  1.406  -76.22807
  1.646  -76.23028
  1.886  -76.23030
  2.196  -76.23030
  2.199  -76.23030
  2.291  -76.23563
  2.501  -76.24244
  2.731  -76.27651
  2.941  -76.28140
  3.131  -76.28471
  3.341  -76.28703
  3.561  -76.28848
  3.761  -76.28931
  3.951  -76.28978
  4.191  -76.29005
  4.421  -76.29020
  4.631  -76.29032
  4.901  -76.29046
  5.211  -76.29059
  5.581  -76.29068
  5.584  -76.29068
  5.675  -76.29938
  5.985  -76.30280
  6.305  -76.30530
  6.615  -76.30741
  6.905  -76.30914
  7.195  -76.31060
  7.485  -76.31196
  7.785  -76.31329
  8.125  -76.31447
  8.455  -76.31543
  8.755  -76.31617
  9.035  -76.31674
  9.325  -76.31720
  9.595  -76.31756
  9.875  -76.31780
 10.165  -76.31798
 10.495  -76.31810
 10.765  -76.31819
 11.065  -76.31825
 11.355  -76.31829
 11.645  -76.31831
 11.925  -76.31833
 12.215  -76.31834
 12.565  -76.31835
 12.569  -76.31835
 12.669  -76.32234
 13.009  -76.32457
 13.359  -76.32603
 13.699  -76.32691
 14.009  -76.32751
 14.339  -76.32789
 14.709  -76.32816
 15.059  -76.32840
 15.379  -76.32860
 15.689  -76.32875
 16.009  -76.32887
 16.329  -76.32895
 16.649  -76.32901
 16.979  -76.32904
 17.369  -76.32907
 17.374  -76.32907
 17.484  -76.33185
 17.894  -76.33294
 18.274  -76.33389
 18.634  -76.33459
 18.994  -76.33509
 19.344  -76.33546
 19.724  -76.33572
 20.124  -76.33592
 20.514  -76.33608
 20.884  -76.33618
 21.254  -76.33627
 21.614  -76.33633
 21.974  -76.33638
 22.344  -76.33641
 22.724  -76.33643
 23.094  -76.33645
 23.444  -76.33647
 23.884  -76.33649
 24.354  -76.33650
 24.864  -76.33651
 25.224  -76.33652
 25.231  -76.33652
 25.371  -76.33655
 25.541  -76.33685
 25.901  -76.33863
 26.251  -76.33950
 26.601  -76.34033
 26.951  -76.34109
 27.431  -76.34178
 27.861  -76.34238
 28.261  -76.34286
 28.701  -76.34323
 29.351  -76.34350
 29.971  -76.34370
 30.471  -76.34383
 30.951  -76.34392
 31.491  -76.34398
 31.931  -76.34403
 32.291  -76.34407
 32.651  -76.34411
 33.041  -76.34414
 33.441  -76.34417
 33.841  -76.34420
 34.251  -76.34423
 34.611  -76.34425
 34.961  -76.34428
 35.331  -76.34429
 35.691  -76.34431
 36.061  -76.34432
 36.551  -76.34432
 36.559  -76.34433
 36.739  -76.34435
 36.929  -76.34499
 37.509  -76.34521
 38.079  -76.34547
 38.669  -76.34571
 39.339  -76.34592
 39.979  -76.34609
 40.529  -76.34622
 41.179  -76.34631
 41.759  -76.34639
 42.329  -76.34644
 42.869  -76.34648
 43.429  -76.34650
 43.999  -76.34652
 44.599  -76.34653
 45.139  -76.34655
 45.679  -76.34657
 46.249  -76.34659
 46.789  -76.34660
 47.339  -76.34661
 47.899  -76.34662
 48.439  -76.34663
 48.450  -76.34663
 48.650  -76.34665
 48.880  -76.34667
 49.120  -76.34670
  };

  \addplot[Green, mark=*, line width=0.5pt, mark size=2pt, mark repeat=3] table {
  0.066  -70.92550
  0.132  -73.55497
  0.198  -74.80562
  0.258  -75.49185
  0.321  -75.88558
  0.383  -76.07463
  0.449  -76.15667
  0.509  -76.19199
  0.572  -76.20372
  0.633  -76.20839
  0.694  -76.21183
  0.756  -76.21563
  0.817  -76.21851
  0.877  -76.22164
  0.941  -76.22380
  1.003  -76.22637
  1.067  -76.22711
  1.127  -76.22761
  1.190  -76.22810
  1.259  -76.22842
  1.322  -76.22807
  1.552  -76.23028
  1.832  -76.23030
  2.142  -76.23030
  2.144  -76.23030
  2.384  -76.31017
  2.614  -76.32496
  2.844  -76.33445
  3.074  -76.33880
  3.274  -76.34074
  3.484  -76.34253
  3.704  -76.34357
  3.914  -76.34347
  4.454  -76.34728
  4.994  -76.34746
  5.534  -76.34760
  6.094  -76.34772
  6.634  -76.34781
  7.174  -76.34788
  7.714  -76.34794
  8.254  -76.34800
  8.784  -76.34805
  9.324  -76.34809
  9.864  -76.34813
 10.404  -76.34818
 10.924  -76.34822
 11.464  -76.34825
 12.004  -76.34829
 12.544  -76.34831
 13.074  -76.34834
 13.784  -76.34836
 13.797  -76.34836
  };
    
  \end{axis}
\end{tikzpicture}
  \end{subfigure}
  \caption{Energy profile along the calculation of \ce{H2O}/cc-pVTZ. One-shot corresponds to a single calculation with the maximum value of $\mathrm{N_g}$ possible, Incremental corresponds to a perfect pairing calculation followed by several restarts including one more orbital on the subspace up to the maximum value of $\mathrm{N_g}$, and Two-step correspond to a perfect pairing calculation followed by a restart with the maximum value of $\mathrm{N_g}$.}
  \label{fig:h2o-OneShot-IncrementalDirect}
\end{figure}

Another key comparison is to assess whether performing a one-shot calculation, where the maximum number of coupled weakly occupied orbitals allowed by the basis set is included from the start, is more efficient than the incremental approach. Unlike incremental calculations, the one-shot method does not rely on previously optimized orbitals and instead starts the optimization from scratch. Since the one-shot calculation is independent of the orbital sorting scheme used to form the subspaces, this comparison is carried out exclusively using the new alternating orbital sorting strategy. The results are presented in Figure~\ref{fig:h2o-OneShot-IncrementalDirect}, which displays two panels: the top panel shows the energy evolution as a function of external iterations, while the bottom panel presents the energy progression as a function of total computational time.

For the one-shot approach, all red marks in Figure~\ref{fig:h2o-OneShot-IncrementalDirect} correspond to calculations where $\mathrm{N_g^{max}}$ is used from the beginning. In contrast, the incremental approach, shown in blue, starts with $\mathrm{N_g} = 1$ and progressively increases the number of coupled weakly occupied orbitals until reaching $\mathrm{N_g^{max}}$. At first glance, the top panel reveals that the incremental approach requires a significantly larger number of external iterations compared to the one-shot calculation. However, this observation alone does not provide a complete picture as not all iterations contribute equally to the total computational cost. Most of the iterations in the incremental approach are substantially faster than those in the one-shot approach, which is evident in the bottom panel, where the time difference between the two approaches is notably reduced. This suggests that while the incremental approach involves more iterations, its computational efficiency per iteration partially offsets the higher iteration count.

To further investigate possible optimizations, we tested an intermediate strategy labeled the two-step approach, which is represented by green marks in Figure~\ref{fig:h2o-OneShot-IncrementalDirect}. In this method, we begin with a perfect pairing calculation ($\mathrm{N_g} = 1$) and then directly transition to $\mathrm{N_g^{max}}$ without going through the intermediate steps of the incremental approach. The results show that the two-step method offers a clear advantage over the one-shot approach in both the number of iterations and the total computational time. This efficiency gain is attributed to the fact that starting with a perfect pairing solution allows the optimization process to begin from a well-defined reference state, reducing the complexity of the full optimization problem when the subspace is expanded to $\mathrm{N_g^{max}}$.

Overall, the findings from Figure~\ref{fig:h2o-OneShot-IncrementalDirect} indicate that, while the one-shot approach remains viable, the incremental and two-step approaches offer practical advantages. The incremental approach leverages previously optimized orbitals, leading to smoother convergence and better numerical stability, whereas the two-step approach provides a computationally efficient balance between accuracy and cost. These results demonstrate that by strategically structuring the orbital sorting and optimization process, significant improvements in computational efficiency can be achieved without compromising the accuracy of the electronic structure description.

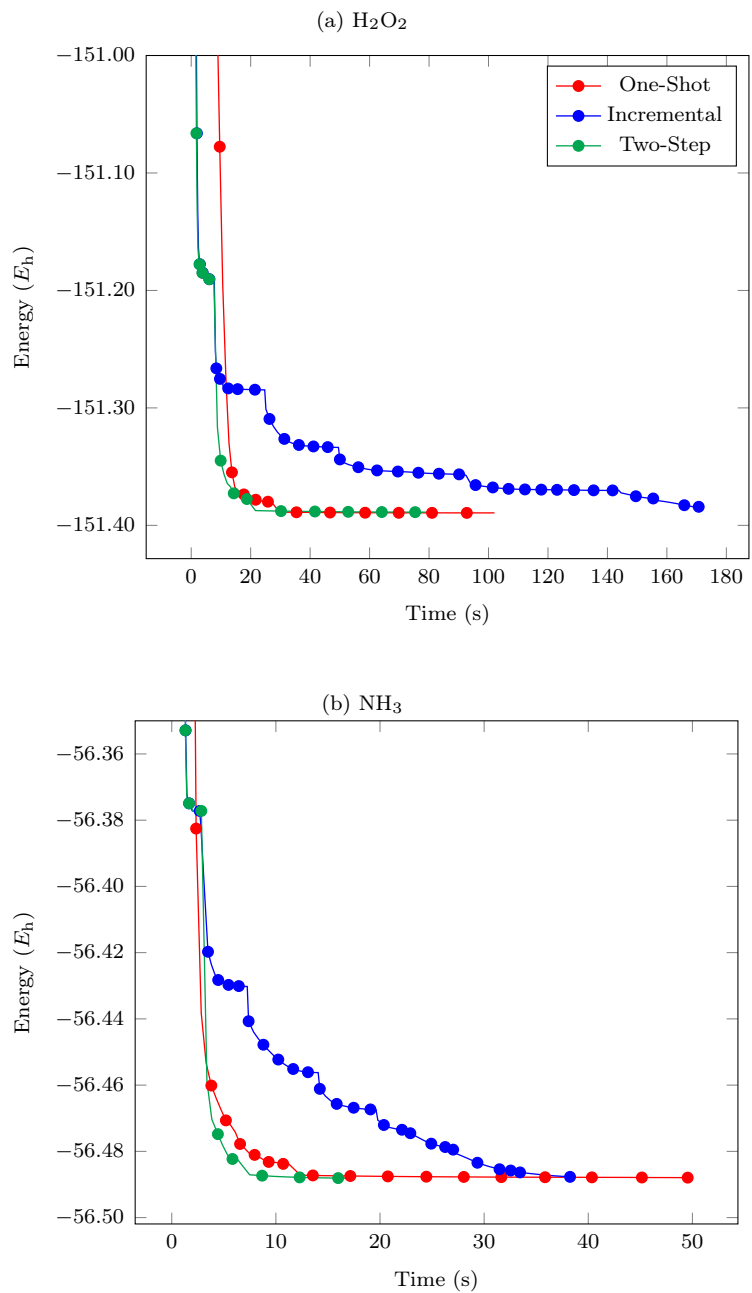
\begin{figure}[htbp]
  \centering
  \begin{subfigure}{0.7\textwidth}
    \caption{\ce{H2O2}}
    \begin{tikzpicture}[font=\scriptsize]
  \begin{axis}[
    xlabel={Time (s)},
    ylabel={Energy ($E_\text{h}$)},
    ymax=-151,
    y tick label style={
        /pgf/number format/.cd,
        fixed,
        fixed zerofill,
        precision=2,
        /tikz/.cd
    },
    width=\textwidth,
  ]
  \addplot[red, mark=*, line width=0.5pt, mark size=2pt, mark repeat=4] table {
  1.300  -141.53274
  2.500  -146.90974
  3.500  -149.32128
  4.500  -150.24349
  5.500  -150.55572
  6.600  -150.71047
  7.600  -150.82987
  8.600  -150.94855
  9.600  -151.07766
 10.600  -151.19502
 11.700  -151.27673
 12.700  -151.32964
 13.700  -151.35486
 14.700  -151.36783
 15.700  -151.36995
 16.700  -151.37202
 17.700  -151.37367
 18.700  -151.37569
 19.700  -151.37703
 20.700  -151.37806
 21.700  -151.37819
 22.700  -151.37925
 23.800  -151.37937
 24.800  -151.37975
 25.800  -151.37986
 26.800  -151.37921
 29.600  -151.38876
 32.500  -151.38881
 35.400  -151.38887
 38.300  -151.38893
 41.100  -151.38898
 43.900  -151.38904
 46.700  -151.38909
 49.500  -151.38914
 52.300  -151.38918
 55.700  -151.38921
 58.500  -151.38924
 61.300  -151.38926
 64.100  -151.38928
 67.000  -151.38930
 69.800  -151.38931
 72.600  -151.38932
 75.400  -151.38933
 78.200  -151.38934
 81.000  -151.38935
 83.900  -151.38936
 86.900  -151.38937
 89.900  -151.38938
 92.700  -151.38938
 95.500  -151.38939
 98.300  -151.38939
102.000  -151.38939
    };
    \addlegendentry{One-Shot}

  \addplot[blue, mark=*, line width=0.5pt, mark size=2pt, mark repeat=4] table {
  0.280  -140.93442
  0.560  -146.48360
  0.800  -148.97191
  1.110  -150.20298
  1.400  -150.75946
  1.660  -150.95750
  1.920  -151.06620
  2.200  -151.13347
  2.440  -151.16374
  2.670  -151.17254
  2.920  -151.17769
  3.150  -151.18105
  3.390  -151.18205
  3.640  -151.18410
  3.880  -151.18495
  4.120  -151.18337
  4.790  -151.19038
  5.460  -151.19042
  6.130  -151.19044
  6.850  -151.19044
  7.760  -151.19045
  8.140  -151.25211
  8.450  -151.26634
  8.750  -151.26924
  9.040  -151.27196
  9.320  -151.27373
  9.670  -151.27530
 10.030  -151.27443
 10.880  -151.28269
 11.660  -151.28309
 12.440  -151.28340
 13.240  -151.28363
 14.030  -151.28379
 14.850  -151.28392
 15.630  -151.28403
 16.630  -151.28415
 18.030  -151.28426
 19.730  -151.28439
 21.430  -151.28451
 23.130  -151.28460
 24.730  -151.28466
 25.120  -151.30105
 26.320  -151.30942
 27.720  -151.31595
 28.920  -151.32037
 30.120  -151.32371
 31.320  -151.32639
 32.520  -151.32847
 33.820  -151.32988
 35.020  -151.33085
 36.220  -151.33155
 37.520  -151.33204
 38.720  -151.33239
 39.920  -151.33264
 41.120  -151.33282
 42.320  -151.33297
 43.520  -151.33311
 44.720  -151.33325
 45.920  -151.33338
 47.120  -151.33350
 48.320  -151.33359
 49.520  -151.33365
 50.010  -151.34388
 51.510  -151.34681
 53.110  -151.34842
 54.710  -151.34953
 56.210  -151.35050
 57.710  -151.35140
 59.210  -151.35219
 60.810  -151.35279
 62.510  -151.35323
 64.210  -151.35357
 66.010  -151.35381
 67.810  -151.35400
 69.510  -151.35417
 71.310  -151.35434
 73.010  -151.35456
 74.710  -151.35486
 76.410  -151.35519
 78.110  -151.35549
 79.810  -151.35572
 81.510  -151.35589
 83.310  -151.35604
 85.010  -151.35618
 86.710  -151.35632
 88.410  -151.35646
 90.110  -151.35657
 91.810  -151.35665
 92.450  -151.35753
 94.050  -151.36469
 95.650  -151.36570
 97.250  -151.36636
 98.750  -151.36688
100.150  -151.36732
101.450  -151.36772
102.750  -151.36809
104.050  -151.36842
105.350  -151.36869
106.750  -151.36891
108.150  -151.36909
109.450  -151.36923
110.950  -151.36933
112.250  -151.36941
113.650  -151.36946
114.950  -151.36951
116.350  -151.36955
117.750  -151.36959
119.050  -151.36962
120.450  -151.36966
121.750  -151.36970
123.050  -151.36975
124.350  -151.36980
125.750  -151.36985
127.150  -151.36991
128.750  -151.36997
130.550  -151.37004
132.150  -151.37010
133.750  -151.37015
135.350  -151.37019
136.950  -151.37022
138.550  -151.37023
140.150  -151.37024
141.750  -151.37024
143.850  -151.37024
144.560  -151.37232
147.060  -151.37373
149.560  -151.37521
152.060  -151.37630
154.460  -151.37703
154.511  -151.37706
155.401  -151.37712
156.361  -151.37871
159.861  -151.37997
163.161  -151.38140
165.861  -151.38284
168.561  -151.38403
168.622  -151.38411
169.602  -151.38415
170.702  -151.38420
  };
    \addlegendentry{Incremental}

  \addplot[Green, mark=*, line width=0.5pt, mark size=2pt, mark repeat=4] table {
  0.240  -140.93442
  0.480  -146.48360
  0.730  -148.97191
  0.970  -150.20298
  1.210  -150.75946
  1.470  -150.95750
  1.750  -151.06620
  2.010  -151.13347
  2.270  -151.16374
  2.510  -151.17254
  2.780  -151.17769
  3.020  -151.18105
  3.270  -151.18205
  3.500  -151.18410
  3.760  -151.18495
  4.010  -151.18337
  4.700  -151.19038
  5.380  -151.19042
  6.050  -151.19044
  6.730  -151.19044
  7.610  -151.19045
  8.810  -151.31641
  9.910  -151.34489
 11.010  -151.35635
 12.110  -151.36422
 13.210  -151.36722
 14.310  -151.37275
 15.410  -151.37413
 16.510  -151.37560
 17.610  -151.37834
 18.710  -151.37757
 21.610  -151.38741
 24.510  -151.38759
 27.310  -151.38774
 30.210  -151.38786
 33.210  -151.38796
 36.010  -151.38805
 38.810  -151.38814
 41.610  -151.38821
 44.410  -151.38828
 47.210  -151.38835
 50.010  -151.38841
 52.810  -151.38847
 55.610  -151.38853
 58.410  -151.38859
 61.310  -151.38863
 64.110  -151.38866
 66.910  -151.38869
 69.710  -151.38872
 72.510  -151.38874
 75.310  -151.38875
 78.210  -151.38876
 82.110  -151.38877
  };
    \addlegendentry{Two-Step}
    
  \end{axis}
\end{tikzpicture}
  \end{subfigure}
  \hfill
  \begin{subfigure}{0.7\textwidth}
    \caption{\ce{NH3}}
    \begin{tikzpicture}[font=\scriptsize]
  \begin{axis}[
    xlabel={Time (s)},
    ylabel={Energy ($E_\text{h}$)},
    ymax=-56.35,
    y tick label style={
        /pgf/number format/.cd,
        fixed,
        fixed zerofill,
        precision=2,
        /tikz/.cd
    },
    width=\textwidth,
  ]
  \addplot[red, mark=*, line width=0.5pt, mark size=2pt, mark repeat=3] table {
  0.470  -50.86408
  0.930  -53.70798
  1.390  -55.51133
  1.880  -56.20031
  2.340  -56.38254
  2.830  -56.43824
  3.300  -56.45331
  3.800  -56.46013
  4.270  -56.46399
  4.740  -56.46749
  5.200  -56.47065
  5.660  -56.47212
  6.110  -56.47426
  6.560  -56.47777
  7.030  -56.47920
  7.480  -56.48069
  7.960  -56.48106
  8.410  -56.48179
  8.850  -56.48257
  9.310  -56.48319
  9.770  -56.48330
 10.230  -56.48345
 10.700  -56.48378
 11.150  -56.48355
 12.350  -56.48718
 13.550  -56.48727
 14.750  -56.48734
 15.950  -56.48739
 17.150  -56.48744
 18.350  -56.48748
 19.550  -56.48752
 20.750  -56.48756
 21.950  -56.48760
 23.150  -56.48763
 24.450  -56.48765
 25.650  -56.48768
 26.850  -56.48770
 28.050  -56.48772
 29.250  -56.48774
 30.450  -56.48776
 31.650  -56.48778
 32.950  -56.48779
 34.250  -56.48780
 35.850  -56.48781
 37.450  -56.48782
 38.650  -56.48783
 40.350  -56.48785
 42.450  -56.48786
 43.750  -56.48787
 45.150  -56.48789
 46.450  -56.48791
 47.850  -56.48792
 49.550  -56.48794
    };

  \addplot[blue, mark=*, line width=0.5pt, mark size=2pt, mark repeat=3] table {
  0.110  -51.63806
  0.220  -53.67571
  0.330  -54.62475
  0.430  -55.08992
  0.540  -55.40852
  0.650  -55.66505
  0.780  -55.87807
  0.890  -56.02761
  0.990  -56.14844
  1.110  -56.23905
  1.220  -56.30875
  1.330  -56.35288
  1.450  -56.37195
  1.560  -56.37502
  1.670  -56.37492
  1.970  -56.37719
  2.280  -56.37720
  2.680  -56.37720
  2.800  -56.38195
  3.160  -56.40236
  3.480  -56.41974
  3.800  -56.42355
  4.130  -56.42646
  4.460  -56.42826
  4.790  -56.42908
  5.120  -56.42951
  5.450  -56.42974
  5.790  -56.42989
  6.110  -56.43000
  6.440  -56.43009
  6.790  -56.43016
  7.250  -56.43022
  7.380  -56.44071
  7.860  -56.44398
  8.330  -56.44606
  8.800  -56.44780
  9.280  -56.44946
  9.750  -56.45095
 10.230  -56.45228
 10.700  -56.45346
 11.180  -56.45443
 11.660  -56.45513
 12.130  -56.45560
 12.610  -56.45592
 13.080  -56.45610
 13.570  -56.45619
 14.060  -56.45622
 14.220  -56.46113
 14.760  -56.46333
 15.280  -56.46478
 15.840  -56.46567
 16.380  -56.46619
 16.920  -56.46655
 17.470  -56.46683
 18.000  -56.46706
 18.540  -56.46723
 19.080  -56.46734
 19.620  -56.46742
 19.790  -56.47057
 20.370  -56.47204
 20.960  -56.47282
 21.540  -56.47322
 22.110  -56.47348
 22.710  -56.47372
 22.723  -56.47373
 22.913  -56.47452
 23.563  -56.47562
 24.273  -56.47681
 24.923  -56.47768
 25.583  -56.47827
 26.243  -56.47868
 26.256  -56.47869
 26.516  -56.47871
 26.756  -56.47945
 27.026  -56.47948
 27.356  -56.48072
 28.356  -56.48204
 29.356  -56.48344
 30.456  -56.48457
 31.456  -56.48535
 31.478  -56.48540
 31.838  -56.48542
 32.178  -56.48577
 32.548  -56.48579
 32.958  -56.48582
 32.990  -56.48582
 33.450  -56.48636
 35.050  -56.48691
 36.650  -56.48736
 38.250  -56.48770
 38.287  -56.48773
  };

  \addplot[Green, mark=*, line width=0.5pt, mark size=2pt, mark repeat=3] table {
  0.120  -51.63806
  0.230  -53.67571
  0.340  -54.62475
  0.450  -55.08992
  0.560  -55.40852
  0.680  -55.66505
  0.790  -55.87807
  0.900  -56.02761
  1.000  -56.14844
  1.110  -56.23905
  1.210  -56.30875
  1.310  -56.35288
  1.430  -56.37195
  1.530  -56.37502
  1.660  -56.37492
  2.030  -56.37719
  2.440  -56.37720
  2.840  -56.37720
  3.380  -56.45896
  3.850  -56.47035
  4.430  -56.47480
  4.880  -56.47814
  5.380  -56.48123
  5.830  -56.48230
  6.280  -56.48231
  7.480  -56.48707
  8.680  -56.48734
  9.880  -56.48756
 11.080  -56.48773
 12.280  -56.48786
 13.480  -56.48795
 14.780  -56.48802
 15.980  -56.48807
 17.580  -56.48811
  };
    
  \end{axis}
\end{tikzpicture}
  \end{subfigure}
  \caption{Profile of the energy along the calculation time for the \ce{H2O2} ($\mathrm{N_g^{max}} = 10$) and \ce{NH3} ($\mathrm{N_g^{max}} = 15$) with the cc-pVTZ bassis set, using the One-Shot, the Incremental and the Two-Step approaches.}
  \label{fig:h2o2+nh3-OneShot-IncrementalDirect}
\end{figure}

Additional validation of the different approaches is presented in Figure~\ref{fig:h2o2+nh3-OneShot-IncrementalDirect}, which examines hydrogen peroxide (\ce{H2O2}) and ammonia (\ce{NH3}) as test cases. These molecules provide a broader perspective on the performance of orbital sorting strategies in chemically distinct systems. Given that our primary objective is to accelerate computations, the discussion focuses on the total computational time required to reach convergence. The top panel of Figure~\ref{fig:h2o2+nh3-OneShot-IncrementalDirect} displays the energy evolution as a function of calculation time for \ce{H2O2}, while the bottom panel presents the corresponding results for \ce{NH3}.

Consistent with previous observations, the incremental approach exhibits a monotonic energy decrease as the subspace size expands, reflecting a smooth and controlled convergence. This behavior is particularly advantageous as it allows the reuse of previously optimized orbitals, leading to improved numerical stability and a reduction in the number of expensive re-optimizations. Notably, this sequential refinement process is only feasible due to the alternating orbital sorting strategy, which preserves the previously converged solution while allowing new orbitals to be incorporated incrementally. Without this sorting scheme, the reordering of orbitals between steps would disrupt the optimization process, increasing the computational cost.

Among the three approaches tested, the two-step strategy, which consists of performing a perfect pairing calculation ($\mathrm{N_g} = 1$) followed by an extended pairing calculation with the maximum number of coupled orbitals ($\mathrm{N_g^{max}}$), emerges as the most efficient scheme. As illustrated in Figure~\ref{fig:h2o2+nh3-OneShot-IncrementalDirect}, this approach consistently achieves the lowest total computation time. The two-step approach leverages an optimized initial reference (perfect pairing) to accelerate convergence when transitioning to the full subspace, thereby outperforming both the incremental and one-shot approaches in terms of efficiency.

The results in Figure~\ref{fig:h2o2+nh3-OneShot-IncrementalDirect} further support the generality of the proposed sorting and optimization strategies. The efficiency gains observed for \ce{H2O2} and \ce{NH3} demonstrate that the benefits of the alternating orbital sorting scheme, as well as the incremental and two-step approaches, extend beyond the specific case of water. These findings confirm that the two-step approach is the optimal computational strategy, making it a robust and scalable method for NOF calculations in correlated molecular systems.

\section{Conclusion}

The introduction of the alternating orbital sorting scheme significantly enhances the computational efficiency and flexibility of the DoNOF program, enabling a more adaptive optimization process that efficiently reuses previously converged solutions. By preserving orbital information across calculations, this approach reduces computational time and improves numerical stability, making it particularly beneficial for systems where computational resources are constrained.

Benchmark calculations on \ce{H2O}, \ce{H2O2}, and \ce{NH3} confirm that the alternating orbital sorting scheme outperforms previous one-shot approaches, achieving faster convergence while maintaining accuracy. Among the tested strategies, the two-step method emerges as the most efficient, leveraging an optimized initial reference (perfect pairing) to accelerate the transition to the full subspace, thereby reducing overall computational cost. These findings demonstrate that the proposed sorting scheme provides an efficient and scalable framework for improving NOF calculations.

Beyond the current study, the alternating orbital sorting scheme opens new opportunities for extending NOFT to larger and more complex systems. The present results highlight how small orbital subspaces can be strategically used to facilitate faster convergence in larger subspaces within the same basis set. Future work will explore extending this approach by incorporating solutions from small basis sets into calculations with larger basis sets, ultimately improving computational efficiency while ensuring an accurate description of both static and dynamic electron correlations.

\section*{Acknowledgments}

All authors acknowledge the Donostia International Physics Center (DIPC) for its support. E. Boutou thanks Polytech Clermont-Ferrand, the Erasmus program, and the Auvergne Rh\^{o}ne-Alpes region for funding her internship. J. F. H. Lew-Yee acknowledges the MCIN program "Severo Ochoa" under reference AEI/CEX2018-000867-S for postdoctoral funding (Ref.: 2023/74). This work has received financial support from the Ministry of Science, Innovation, and Universities (MCIU) under project PID2021-126714NB-I00 and from the Basque Government (Eusko Jaurlaritza, IT1584-22). The authors also thank SGIker (UPV/EHU/ERDF, EU) for providing technical and human support, as well as for the allocation of computational resources through the Scientific Computing Service.

\appendix
\renewcommand{\theequation}{A\arabic{equation}} 
\setcounter{equation}{0}  

\section*{Appendix}

Consider a mixed singlet state of N spin-paired electrons. The Global Natural Orbital Functional (GNOF) follows an electron-pairing approach, where the orbital space $\Omega$ is divided into N/2 mutually disjoint subspaces $\Omega_g$. Each orbital belongs exclusively to a single subspace $\Omega_g$, which consists of one strongly double-occupied orbital $g$ and N$_{g}$ weakly double-occupied orbitals. This partitioning ensures that the sum of the occupation numbers within each subspace is exactly two, while maintaining that the trace of the one-particle reduced density matrix correctly corresponds to the total number of electrons.

The reconstruction functional for the two-particle reduced density matrix in terms of the occupation numbers leads to GNOF, as illustrated by the equation:
\begin{equation}
E_\text{el}=E^\text{intra}+E_\text{HF}^\text{inter}+E_\text{sta}^\text{inter}+E_\text{dyn}^\text{inter} \label{gnof}
\end{equation}
\noindent The intra-pair component is formed by summing the energies of electron pairs, specifically,
\begin{equation}
E^\text{intra}=\sum\limits _{g=1}^{\mathrm{N}/2} \left[ \,\, 2 \sum\limits _{p\in\Omega_{g}}n_{p}H_{pp} + \sum\limits _{p,q\in\Omega_{g}} \Pi(n_q,n_p) L_{pq} \right]
\label{Eintra}
\end{equation}
\noindent where $H_{pp}$ are the diagonal one-electron matrix elements of the kinetic energy and external potential operators, whereas $L_{pq}=\left\langle pp|qq\right\rangle$ are the exchange-time-inversion integrals. The matrix elements $\Pi(n_q,n_p) = c(n_q)c(n_p)$, where $c(n_p)$ is defined by the square root of the occupation numbers according to the following rule:
\begin{equation}
    c(n_p) = \left.
  \begin{cases}
    \phantom{+}\sqrt{n_p}, & p \leq \mathrm{N}/2\\
    -\sqrt{n_p}, & p > \mathrm{N}/2 \\
  \end{cases}
  \right. \>\>
  \label{eq:PNOF5-roots}
\end{equation}
that is, the phase factor of $c(n_p)$ is chosen to be $+1$ for the strongly occupied orbital of a given subspace $\Omega_g$, and $-1$ otherwise. The inter-subspace Hartree-Fock (HF) term is 
\begin{equation}
E_\text{HF}^\text{inter}=\sum\limits _{p,q}\,'\,n_{q}n_{p}\left(2J_{pq}-K_{pq}\right)\label{ehf}
\end{equation}
where $J_{pq}=\left\langle pq|pq\right\rangle$ and $K_{pq}=\left\langle pq|qp\right\rangle $ are the Coulomb and exchange integrals, respectively. The prime in the summation indicates that only the inter-subspace terms are taken into account. The inter-subspace static component is written as 
\begin{equation}
E_\text{sta}^\text{inter} = - \sum\limits _{p,q}\,'\,\left(1-\delta_{q\Omega^{b}}\delta_{p\Omega^{b}}\right) \Phi_{q}\Phi_{p}L_{pq}
\label{esta}
\end{equation}
where $\Phi_{p}=\sqrt{n_{p}h_{p}}$ with the hole $h_{p}=1-n_{p}$. Finally, the inter-subspace dynamic energy is
\begin{equation}
E_\text{dyn}^\text{inter}=\sum\limits _{p,q}\,'\,\left(1-\delta_{q\Omega^{b}}\delta_{p\Omega^{b}}\right) \left[\Pi(n_{q}^{d},n_{p}^{d})+n_{q}^{d}n_{p}^{d}\right] L_{pq}
\label{edyn}
\end{equation}
In Eqs. (\ref{esta}) and (\ref{edyn}), $\Omega^{b}$ denotes the subspace composed of orbitals below the level $\mathrm{N}/2$, whereas $n_{p}^{d}$ is the dynamic part of the occupation number $n_{p}$ in accordance with the Pulay\textquoteright s criterion that establishes an occupancy deviation of approximately 0.01 with respect to 1 or 0 for a natural orbital to contribute to the dynamic correlation.


\end{document}